\documentstyle[aps,prb,floats,eqsecnum,epsf]{revtex}
\begin{document}
\tighten
\draft
\def\figurename{Fig.}
%----------------------------------------------------------------
\title{Boundary Integral Method for Stationary States of Two-Dimensional
  Quantum Systems} 

\author{Ioan Kosztin and Klaus Schulten}
\address{Department of Physics, University of Illinois at Urbana-Champaign,\\
1110 West Green Street, Urbana, Illinois 61801}
\date{January 3, 1997}
\maketitle
%----------------------------------------------------------------
\begin{abstract}
  The {\em boundary integral method\/} for calculating the stationary states
  of a quantum particle in nano-devices and quantum billiards is presented in
  detail at an elementary level. According to the method, wave functions
  inside the domain of the device or billiard are expressed in terms of line
  integrals of the wave function and its normal derivative along the
  domain's boundary; the respective energy eigenvalues are obtained as the
  roots of Fredholm determinants.  Numerical implementations of the
  method are described and applied to determine the energy level statistics
  of billiards with circular and stadium shapes and demonstrate the quantum
  mechanical characteristics of chaotic motion.  The treatment of other
  examples as well as the advantages and limitations of the boundary integral
  method are discussed.
\end{abstract}
\pacs{E-print: {\tt physics/9702022}}

%----------------------------------------------------------------------------
\typeout{Section: Introduction}
\section{Introduction}
\label{sec:intro}

Recent advances in {\em nanotechnology}, based on advanced crystal growth
and lithographic techniques, have opened an avenue to fabricate very
small and clean electronic devices, known as {\em nano-devices}\cite{buot93}.
The charge carriers (electrons) in such devices, through gate voltages, are
confined to one or two spatial dimensions. At very low temperatures, the
spatial extent of the systems along the direction of confinement is
comparable to the Fermi wavelength of the electrons. {\em Quantum dots\/}
and {\em quantum wires\/} are examples of quasi zero- and one-dimensional
nano-devices in which confinement of the electrons occur along all three and
along two spatial directions, respectively, while in the inversion layer of
narrow-gap semiconductor heterostructures the electrons are confined along
the direction perpendicular to the layer. Quantum dots are relevant in the
study of the {\em Coulomb blockade\/} phenomena\cite{tinkham96}, while
quantum wires are experimental realizations of so-called {\em Luttinger
  liquids}\cite{luttinger_liquid}.

The motion of the electrons in a clean two-dimensional nano-device is
ballistic, i.e., the electrons are scattered mainly by the device boundaries
and not by impurities. The device boundaries, due to high precision
lithography, may have arbitrary shapes and are very sharp, i.e., the
electrical potential changes abruptly on atomic scales. As a result, the
behavior of such two-dimensional nano-devices, which exhibit quantum
confinement in one direction and free motion of the electrons in a finite
two-dimensional domain of sub-micron size, is governed by single-electron
(particle) physics, and can be described theoretically by solving the
corresponding Schr\"odinger wave equation. Such nano-devices can be
considered as quantum mechanical analogue of {\em classical billiard\/}
systems\cite{berry81} in which point like particles bounce inside a
two-dimensional (2D) region ${\cal D}$ delimited by the contour $\Gamma$.
An idealized quantum billiard confines a quantum particle inside a 2D infinite
potential well; the shape of the infinite well being determined by $\Gamma$.

Quantum billiards represent models of nano-devices which play an important
role in modern semiconductor industry\cite{buot93}. The experimental
study, via STM techniques, of quantum billiards provides a new
testing-ground for the predictions of quantum mechanics\cite{buot93}. The
study of quantum billiards allows one to investigate also the quantum
signatures of classical chaos. It is known that non-integrable classical
systems are chaotic, i.e., the phase space trajectory of the system exhibits
exponential sensitivity to the initial conditions. In the case of billiards,
the chaotic behavior is caused by the irregularities of the boundary and not
by the complexity of the interaction in the system (e.g., scattering of the
particle from randomly distributed impurities).  Since the concept of
``phase space trajectory'' loses its meaning in quantum mechanics, one can
naturally ask oneself what is the quantum mechanical analogue of (classical)
chaos, or more precisely, is there any detectable difference between the
behavior of a quantum system with chaotic- and non-chaotic classical limit,
respectively.  

The answer to these questions should be sought in the characteristics of the
fluctuations of the energy levels of the quantum billiard
systems\cite{mcdonald79,bohigas84a}. Thus, in order to study the physical
properties of quantum billiards one needs to find first the corresponding
energy spectrum by solving the time independent Schr\"odinger equation

\begin{equation}
  \label{eq:i1}
  \hat{H}\psi_n(\bbox{r}) \;\equiv\; \left[-\frac{\hbar^2}{2m}\nabla^2 +
    V(\bbox{r})\right] \psi_n(\bbox{r})
  \;=\; E_n\psi_n(\bbox{r})\;,
\end{equation}
where $\hat{H}$ is the Hamiltonian of the system, $V(\bbox{r})$ is the
potential energy, and $\psi_n$ is the eigenfunction corresponding to the
energy eigenvalue $E_n$. In general, in (\ref{eq:i1}) the potential
$V(\bbox{r})$ does not contain the term corresponding to the infinite
potential well; the effect of the later is reflected by the ``hard-wall''
(i.e., Dirichlet) boundary conditions at the billiard boundary. The spectrum
is discrete and the distribution of the energy levels $E_n$ is determined by
the form of the potential and by the boundary conditions.

Eq.(\ref{eq:i1}) can be solved analytically only for very few special cases,
when the system is integrable, i.e., when there exists, besides the energy,
a second conserved physical quantity. Such examples, like a quantum particle
in a rectangular or circular infinite potential well, are discussed in most
of the quantum mechanics textbooks\cite{landau-qm} and in some recent
publications\cite{robinett}, as well. However, for a generic quantum
billiard the energy spectrum can be determined only numerically, and the
description of such numerical methods lacks in all widely used quantum
mechanics textbooks.

The purpose of the present article is to fill this gap by providing the
reader with a self-contained and practical introduction to a powerful
numerical method, known as the {\em Boundary Integral Method\/} (BIM), for
calculating the energy levels of a 2D quantum system, e.g., a quantum
billiard. While the BIM, sometimes also referred to as the {\em Boundary
Element Method} (BEM), has been extensively used for many years for solving
different engineering problems\cite{chen-zhou,banerjee,kitahara}, its
application for calculating energy spectra of quantum billiards has emerged
only recently\cite{riddell,amini,berry-wilkinson,boasman,sieber}.

Before we embark on our presentation of the BIM for calculating energy
spectra of 2D quantum systems, let us first mention a few other frequently
used numerical methods in the same context.

Essentially all numerical methods devised to solve the single particle
Schr\"odinger equation (\ref{eq:i1}) can be classified in two groups. The
methods belonging to the first group assume that one readily knows a
complete set of orthonormal functions $\left\{\phi_m(\bbox{r})\right\}$
which obey the desired boundary conditions along the billiard boundary. By
expanding the unknown energy eigenfunctions

\begin{equation}
  \label{eq:i2}
  \psi_n(\bbox{r}) \;=\; \sum_m c_{nm}\phi_m(\bbox{r})\;,
\end{equation}
the Schr\"odinger equation is converted into the familiar system of
homogeneous linear equations for the coefficients $c_{nm}$
\begin{equation}
\sum_m\left(H_{nm}-E_n\delta_{nm}\right)c_{nm} \;=\; 0\;,
  \label{eq:i1a}
\end{equation}
Here $\delta_{nm}$ is the Kronecker-delta (equal to $1$ for $n=m$ and zero
otherwise), and the matrix elements of the Hamiltonian are 
\begin{equation}
  \label{eq:i4}
  H_{nm} \;=\; \int\!\!d\bbox{r}\; \phi_n^*(\bbox{r})\hat{H}\; \phi_m(\bbox{r})\;.
\end{equation}
Equation (\ref{eq:i1a}) admits non-trivial solutions (energy eigenstates or
stationary states) only for those values of $E_n$ (the energy eigenvalues)
which satisfy the condition
\begin{equation}
  \label{eq:i3}
  \text{det}\left|H_{nm}-E_n\delta_{nm}\right| \;=\; 0\;.
\end{equation}
This condition can be employed to determine the $E_n$'s.

When the billiard boundary is irregular, in general, it is impossible to
find analytical expressions for the functions $\phi_m(\bbox{r})$ and,
therefore, the method as described fails. However, in this case one can
overcome the previously mentioned difficulty by either performing a
coordinate transformation which renders the boundary highly regular, or by
extending the system, fitting the billiard inside a rectangle or circle
along which the Dirichlet boundary conditions apply. Now a complete set of
orthonormal functions can be easily found, but the price one pays in both cases
is that the corresponding Hamiltonian becomes more complicated: in the
former case the simple form of the kinetic energy is
altered\cite{nakamura88} while in the latter case the potential energy is
modified\cite{parmley96}, i.e., $V(\bbox{r})=0$ inside ${\cal D}$ and
$V(\bbox{r})=\infty$ (in practice a suitably chosen large value) outside
${\cal D}$.

The second class of numerical methods intended to calculate billiard spectra
regard Eq.(\ref{eq:i1}) as a partial differential equation for which the
general solution is formally given by (\ref{eq:i2}) for some conveniently
chosen basis functions $\phi_m(\bbox{r})$. The energy eigenfunctions and
eigenvalues are determined by requiring the general solution (\ref{eq:i2})
to obey the Dirichlet boundary conditions along $\partial{\cal D}$. Of
course, the boundary conditions can be met only for particular values of the
energy, i.e., the energy eigenvalues. Heller\cite{heller89} used this method
choosing as the basis functions plane waves, while a more general
and systematic implementation of this method in plane polar coordinates is
described by Schmit\cite{schmit89}.

The BIM is an efficient alternative to the above mentioned two classes of
methods for solving numerically the Schr\"odinger equation. We shall
consider its application only for two-dimensional systems. The BIM will
allow us to study the quantum analogue of classical chaotic systems and
reveal that chaotic behavior is reflected in the spacing of the energy
eigenvalues $E_n$. For this purpose, the BIM is formulated in
Sec.~\ref{sec:bim} and is applied, in Sec.~\ref{sec:billiard}, to the
spectra of circular, stadium and generalized stadium billiards. In
Sec.~\ref{sec:examples} we discuss further examples to which the BIM can be
applied. In Sec.~\ref{sec:conc} we present concluding remarks.

%--------------------------------------------------------------------------

\typeout{Section: The Boundary Integral Method}
\section{The Boundary Integral Method}
\label{sec:bim}

Consider a quantum particle of mass $m$ moving in  a finite,  simply connected
region ${\cal D}$, experiencing the potential
$V(\bbox{r})$ and being governed  by the Hamiltonian 
\begin{equation}
\hat{H} \;=\; -\frac{\hbar^2}{2m}\nabla^2 + V(\bbox{r})\;.
  \label{eq:bim1}
\end{equation}
The energy spectrum of the particle can be determined from the
time-independent Schr\"odinger equation (\ref{eq:i1}) together with the 
boundary conditions for the wave functions $\psi_n(\bbox{r})$ specified on 
a closed curve $\Gamma=\partial{\cal D}$ which delimits the region ${\cal D}$.

The Schr\"odinger equation (\ref{eq:i1})  is an implicit equation for $E_n$ and 
$\psi_n(\bbox{r})$.  This differential equation can be replaced by an implicit integral 
equation which can also serve to determine $E_n$ and  $\psi_n(\bbox{r})$.  
For this purpose, one introduces the 
{\em Green's function}, $G(\bbox{r},\bbox{r'};E)$ of the operator
$E-\hat{H}$,  defined as the solution of 
\begin{equation}
  [E-\hat{H}(\bbox{r})]\,G(\bbox{r},\bbox{r'};E) \;=\;
  \delta(\bbox{r}-\bbox{r'}) \;,
  \label{eq:bim3}
\end{equation}
where $\delta(\bbox{r}-\bbox{r'})$ is the two-dimensional $\delta$-function, $E$ is a complex variable , and  $\bbox{r}$,  $\bbox{r'}$ are arbitrary points in ${\cal D}$.
Multiplying Eq.(\ref{eq:i1}) by $G(\bbox{r},\bbox{r'};E)$,
Eq.(\ref{eq:bim3}) by $\psi_n(\bbox{r})$, and adding the resulting equations
yield 
\begin{equation}
\psi_n(\bbox{r})\delta(\bbox{r}-\bbox{r'}) + \left(E_n-E\right)
\psi_n(\bbox{r}) G(\bbox{r},\bbox{r'};E) \;=\;
G(\bbox{r},\bbox{r'};E)\hat{H}\psi_n(\bbox{r}) -
\psi_n(\bbox{r})\hat{H}G(\bbox{r},\bbox{r'};E) \;.
  \label{eq:bim4}
\end{equation}

We consider now  Eq.(\ref{eq:bim4}) for $E=E_n$.  In this case  the second term on the LHS
vanishes, provided that $G$ is finite (i.e., has no poles) at $E_n$. A
necessary (but not sufficient) condition is 
that $G$ does not obey the same boundary conditions as  $\psi_n$. Inserting the Hamiltonian
(\ref{eq:bim1}) in the RHS of Eq.(\ref{eq:bim4}) eliminates the terms containing the potential energy $V(\bbox{r})$ and Eq.(\ref{eq:bim4}) becomes
\begin{equation}
\psi_n(\bbox{r})\delta(\bbox{r}-\bbox{r'}) \;=\;
\frac{\hbar^2}{2m}
\left[\psi_n(\bbox{r})\nabla^2G\left(\bbox{r},\bbox{r'};E_n\right) - %
G\left(\bbox{r},\bbox{r'};E_n\right)\nabla^2 \psi_n(\bbox{r})\right]\;.
  \label{eq:bim5}
\end{equation}
Recalling the identity $u\nabla^2v = \nabla(u\nabla v)-\nabla u \nabla
v$, valid for any differentiable functions $u(\bbox{r})$ and $v(\bbox{r})$,
the RHS of the above equation can be written as a divergence
\begin{equation}
\psi_n(\bbox{r})\delta(\bbox{r}-\bbox{r'}) \;=\;
\frac{\hbar^2}{2m}
\nabla\cdot
\left[\psi_n(\bbox{r})\nabla G\left(\bbox{r},\bbox{r'};E_n\right) - %
G\left(\bbox{r},\bbox{r'};E_n\right)\nabla \psi_n(\bbox{r})\right]\;.
  \label{eq:bim6}
\end{equation}
Integration with respect to $\bbox{r}$ over the domain ${\cal D}$ yields, on
the LHS, $\psi_n(\bbox{r'})$ since $\bbox{r'}\in{\cal D}$; 
% according to Stoke's theorem~\cite{stokes},
applying Green's formula\cite{green}, the integral on the RHS can be
expressed as a line integral along $\Gamma=\partial{\cal D}$ and
Eq.(\ref{eq:bim6}) becomes
\begin{equation}
\psi_n(\bbox{r'}) \;=\; \frac{\hbar^2}{2m}\oint_{\Gamma}\!ds(\bbox{r}) 
\left[\psi_n(\bbox{r})\partial_{\nu}G\left(\bbox{r},\bbox{r'};E_n\right) - %
G\left(\bbox{r},\bbox{r'};E_n\right)%
\partial_{\nu}\psi_n\left(\bbox{r}\right)\right]\;.
  \label{eq:bim7}
\end{equation}
Here $ds(\bbox{r})$ is the infinitesimal arc length along $\Gamma$ at
$\bbox{r}\in\Gamma$, and the normal derivative $\partial_{\nu}$ is defined
through
\begin{equation}
\partial_{\nu} \;\equiv\; \bbox{\nu}(\bbox{r})\cdot\nabla_{\bbox{r}}\;,
  \label{eq:bim8}
\end{equation}
with $\bbox{\nu}(\bbox{r})$ representing the exterior normal unit vector to
$\Gamma$ at $\bbox{r}\in\Gamma$.  This is the desired integral equation
which, for nano-devices and quantum billiards, provides a simpler avenue to
$E_n$ and $\psi_n(\bbox{r})$ than the Schr\"odinger equation (\ref{eq:i1}).
Note that Eq.~(\ref{eq:bim7}) does not exhibit an explicit dependence on the
potential function $V(\bbox{r})$; the effect of the latter is incorporated
entirely in the Green's function $G(\bbox{r},\bbox{r'};E)$.

The eigenvalues $E_n$ can be obtained by noting that existence of solutions
$\psi_n(\bbox{r})$ implies conditions of the type (\ref{eq:i3}).  We will
adopt a similar strategy for Eq.~(\ref{eq:bim7}) and consider for this
purpose the limit $\bbox{r'}\, \in \, \Gamma$. In this limit Eq.(\ref{eq:bim7})
becomes an implicit equation for $\psi_n(\bbox{r})$ confined solely to the
boundary $\Gamma$ such that a condition like (\ref{eq:i3}) can be postulated
and exploited to determine $E_n$.

\begin{figure}[ht]
  \centerline{\epsfxsize=4in\epsffile{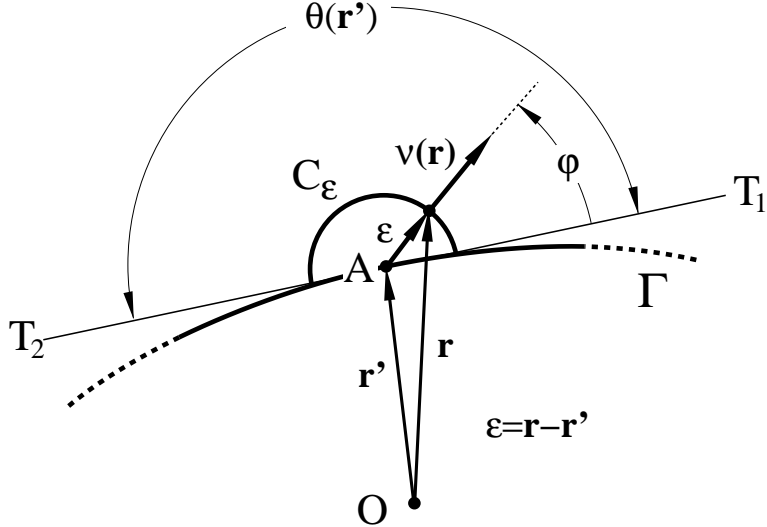}}
  \caption{Geometry of the boundary \protect{$\Gamma_{\varepsilon}$} in the
    vicinity of the point A (position vector {\it\bf r'}) where
    the Green's function is singular.}
  \label{fig:singular}
\end{figure}

The limit $\bbox{r'}\, \in \, \Gamma$ in (\ref{eq:i3}) is not trivial since
both the Green's function and its normal derivative are singular at
$\bbox{r}=\bbox{r'}$. However, these singularities are integrable in the
sense of Cauchy's principal value.  To demonstrate this we carry out the
integration in (\ref{eq:i3}) along a slightly altered contour
$\Gamma_{\varepsilon}$ which avoids the singularity and then let the altered
contour approach $\Gamma$ continuously.  For this purpose we define
$\Gamma_{\varepsilon}=\tilde{\Gamma}_{\varepsilon}\cup C_{\varepsilon}$,
where $\tilde{\Gamma}_{\varepsilon}$ coincides with $\Gamma$, except for a
portion of arc-length $2\varepsilon$ centered about $\bbox{r'}$;
$C_{\varepsilon}$ is a circular arc with center at $\bbox{r'}$ and radius
$\varepsilon$ as shown in Fig.~\ref{fig:singular}, where $\bbox{r'}$
lies inside the region delimited by $\Gamma_{\varepsilon}$.  We consider
then the integral in (\ref{eq:i3}) for $lim_{\varepsilon\rightarrow
  0^+}\Gamma_{\varepsilon}=\Gamma$.

For  $\Gamma_{\varepsilon}$ the integral has two contributions corresponding to
$\tilde{\Gamma}_{\varepsilon}$ and $C_{\varepsilon}$.
The integration along $\tilde{\Gamma}_{\varepsilon}$ in
the $\varepsilon\rightarrow 0^+$ limit is, by definition, Cauchy's principal
value integral along the original contour $\Gamma$. We denote the integral as
\begin{equation}
  \label{eq:bim8a}
  \lim_{\varepsilon\rightarrow 0} \int_{\Gamma_{\varepsilon}}\!ds(\bbox{r}) \ldots
  \equiv {\cal P}\oint_{\Gamma}\!ds(\bbox{r})\ldots\;.
\end{equation}
The contribution due to the integral along $C_{\varepsilon}$ depends on the
type of singularity of the Green's function at $\bbox{r}=\bbox{r'}$.   The integral can be calculated  as shown in  Appendix~\ref{sec:singular-int}.   One obtains
\begin{equation}
  \label{eq:bim9}
  lim_{\varepsilon\rightarrow 0} \frac{\hbar^2}{2m}\int_{C_{\varepsilon}}\!
  ds(\bbox{r})
  \left[\psi_n(\bbox{r})\partial_{\nu}G\left(\bbox{r},\bbox{r'};E_n\right)%
  -G\left(\bbox{r},\bbox{r'};E_n\right)\partial_{\nu}\psi_n(\bbox{r})\right]%
  \;=\; \frac{1}{2}\psi_n(\bbox{r'})\;.
\end{equation}
In the derivation of this formula we have implicitly assumed that there is a
unique tangent to $\Gamma$ at $\bbox{r'}$, i.e., that the angle
$\theta(\bbox{r}')$ in Fig.~\ref{fig:singular} is equal to $\pi$. Otherwise,
according to Eq.(\ref{eq:ap2-5}) in Appendix~\ref{sec:singular-int}, the RHS
of (\ref{eq:bim9}) must be replaced by
$\left(\theta(\bbox{r'})/2\pi\right)\psi_n(\bbox{r'})$, where
$\theta(\bbox{r'})$ is the exterior angle between the two tangents to
$\Gamma$ at $\bbox{r'}$.

Altogether, one obtains for $\psi_n(\bbox{r'})$ , $\bbox{r'} \, \in \, \Gamma$ the integral equation
\begin{equation}
  \label{eq:bim10}
  \psi_n(\bbox{r'}) \;=\; \frac{\hbar^2}{m} {\cal P}\oint_{\Gamma}\!
  ds(\bbox{r}) \left[\psi_n(\bbox{r})\partial_{\nu}% 
  G\left(\bbox{r},\bbox{r'};E_n\right)-G\left(\bbox{r},\bbox{r'};E_n\right)%
  \partial_{\nu}\psi_n(\bbox{r})\right]
\end{equation}
where one still  needs  to specify the boundary
condition on  $\Gamma$ which involves $\psi_n$ and/or its normal
derivative $\partial_{\nu}\psi_n$. The boundary
condition is expressed as  a linear functional relation
\begin{equation}
  {\cal F}\left[\psi_n(\bbox{r}),\partial_{\nu}\psi_n(\bbox{r})\right] \;=\;
  0\;, 
  \qquad \bbox{r}\in\Gamma\;.
  \label{eq:bim11}
\end{equation}
The actual form of the functional ${\cal F}$ depends on the physical problem
at hand, but not on the contour $\Gamma$. The energy eigenvalues $E_n$ are
determined by requiring that Eqs.(\ref{eq:bim10}) and (\ref{eq:bim11}) admit
nontrivial solutions for $\psi_n$. This condition leads us to an equation
involving functional (Fredholm) determinants of the type (\ref{eq:i3}) which
need to be solved by numerical means. Once $E_n$ and the corresponding
$\psi_n$ and $\partial_{\nu}\psi_n$ on $\Gamma$ are determined, the
eigenfunction inside the domain ${\cal D}$ can be calculated using
Eq.(\ref{eq:bim7}).

Below we will demonstrate the application of the method outlined which is
referred to as the {\sl Boundary Integral Method} (BIM).  The method is
practical whenever (i) a Green's function $G$ is available analytically and
(ii) the boundary condition (\ref{eq:bim11}) is fairly simple; the method
applies to $\Gamma$ of arbitrary shape.

\typeout{Section: Billiard Spectra via BIM}
\section{Billiard Spectra via BIM}
\label{sec:billiard}

Inside a billiard a particle moves freely, i.e., $V\, \equiv \, 0$ in
(\ref{eq:bim1}).  The Green's function defined through (\ref{eq:bim3}) is
well known in this case and is given by
\begin{equation}
  G\left(\bbox{r},\bbox{r'};E\right) \;= \;  -\frac{im}{2\hbar^2}
  H_0^{(1)}\left(k\left|\bbox{r}-\bbox{r'}\right|\right) \;,
  \label{eq:b1}
\end{equation}
as shown in Appendix~\ref{sec:free-particle}. Here $k=\sqrt{2mE}/\hbar$ is
the so-called wave vector; the index $n$ is dropped since we focus in the
following on a single eigenstate.  We will also use the notation
$G\left(\bbox{r},\bbox{r'};k\right)$ for the Green's function.  Since the
particle is confined to the billiard, its wave function $\psi\equiv \psi_n$
must vanish along $\Gamma$ and the boundary condition (\ref{eq:bim11}) takes
the form
\begin{equation}
  \psi(\bbox{r})\;=\;0\;,\quad
  \partial_{\nu}\psi(\bbox{r})\;=\;\text{arbitrary}\;, \qquad
  \forall\:\bbox{r}\in\Gamma\;. 
  \label{eq:b2}
\end{equation}
Inserting (\ref{eq:b2}) in the the integral equation (\ref{eq:bim10}) leads to
\begin{equation}
{\cal P}\oint_{\Gamma}\!ds(\bbox{r}) G\left(\bbox{r},\bbox{r'};E\right)
\partial_{\nu}\psi(\bbox{r}) \;=\; 0\;.
  \label{eq:b3}
\end{equation}
This integral equation admits non-trivial solutions only if the
corresponding Fredholm (functional) determinant vanishes, i.e., for
\begin{equation}
\text{det}\left[G\left(\bbox{r},\bbox{r'};E\right)\right] \;=\; 0\;,
  \label{eq:b4}
\end{equation}
a condition  which allows one to determine the energies $E_n$. 

Even though the analytical expression of the Green's function $G$ is known,
the Fredholm determinant (\ref{eq:b4}) is difficult to evaluate for
arbitrary billiard boundaries $\Gamma$. Below we describe more practical
schemes for solving the integral equation (\ref{eq:b3}).

\typeout{Subsection: Methods for Solving the BIE}
\subsection{Methods for Solving the BIE}
\label{sec:methods}

There are basically three different methods for solving the BIE
(\ref{eq:bim10}) for the billiard problem. Before presenting these methods,
let us first parameterize the billiard boundary $\Gamma$ through the arc
length $s\in[0,{\cal L}]$, where ${\cal L}$ is the length of the billiard
boundary $\Gamma$. Thus, the position of each point $\bbox{r}\in\Gamma$ is
uniquely determined by $s$ through the function $\bbox{r}=\bbox{r}(s)$. It is 
convenient to introduce the notation
\begin{equation}
  u(s) \equiv u_n(\bbox{r}(s)) \equiv \partial_{\nu}\psi_n(\bbox{r})\;.
  \label{eq:b5}
\end{equation}
The BIE (\ref{eq:b3}) can be recast then as
\begin{equation}
  \int_0^{\cal L}\!ds \; G\left(s,s';k\right) u(s) \;=\; 0,
  \label{eq:b6}
\end{equation}
where, for brevity, we have dropped the index $n$ which labels the
eigenstates.
 
\typeout{--- Method I.}
{\bf Method I.}\hspace{2ex} The most obvious (but not necessarily the most
convenient) method of solution relies on the observation that both
the wave function and its normal derivative (i.e., $u(s)$) are single-valued
functions and, therefore, $u(s)$ must be a periodic function of $s$ with period
${\cal L}$. Hence, (\ref{eq:b6}) can be expressed as a Fourier
series  
\begin{mathletters}
\begin{equation}
    u(s) \;=\; \sum_{j=-\infty}^{\infty} u_j \exp(iK_js)\;,
    \label{eq:b7a}
  \end{equation}
where
\begin{equation}
  K_j \;\equiv\; \frac{2\pi}{{\cal L}}j\;,\qquad j=0,\pm 1,\pm 2,\ldots
  \label{eq:b7c}
\end{equation}
and where $u_j$ is the Fourier transform of $u(s)$
  \begin{equation}
    u_j \;=\; \frac{1}{{\cal L}}\int_0^{\cal L}\!ds\, u(s)\exp(-iK_js)\;.
    \label{eq:b7b}
  \end{equation}
\end{mathletters}

By taking the Fourier transform of Eq.(\ref{eq:b6}) with respect to $s'$ and
 using (\ref{eq:b7a}) one obtains the system  of linear equations
\begin{equation}
 \sum_{j=-\infty}^{\infty} A_{ij}(k)\, u_j \;=\; 0\;,
  \label{eq:b8}
\end{equation}
where 
\begin{equation}
  A_{ij}(k) \;=\; \frac{1}{{\cal L}}\int_0^{\cal L}\!ds\int_0^{\cal L}\!ds'
  G\left(s,s';k\right) \exp\left[-i(K_is' - K_js)\right]\;.
  \label{eq:b8a}
\end{equation}
Here the information about the
billiard boundary $\Gamma$ is contained in the $s$- and $s'$-dependence 
of the Green's function through $\bbox{r}=\bbox{r}(s)$ and
$\bbox{r'}=\bbox{r}(s')$.   The energy
eigenvalues, expressed through $k$, are  the solutions of the
equation 
\begin{equation}
\text{det}\left[A_{ij}(k)\right] \;=\; 0
  \label{eq:b8c}
\end{equation}
which must hold in order to render (\ref{eq:b8}) solvable.

For an arbitrary $\Gamma$ one cannot solve Eq.(\ref{eq:b8c}) exactly.
However, approximate energy eigenvalues can be obtained by truncating the
infinite system of linear equations (\ref{eq:b8}) at some suitably chosen
wave vector $K_c$. The truncation implies that the Fourier components of
$u(s)$ which correspond to $|K_j|>K_c$ are set equal to zero in
(\ref{eq:b7a}). In this case the relevant part of the matrix $A_{ij}$
becomes finite and the corresponding determinant can be calculated
numerically. The drawback of the truncation is that the resulting energy
eigenvalues expressed through $k$ are accurate only as long as $k\alt K_c$.
If one seeks to describe energy levels with larger $k$-values one needs to
increase $K_c$ which, however, leads to an increased computational effort,
the latter increasing rapidly with the dimension of the matrix $A_{ij}$.

The calculation of  the matrix elements $A_{ij}$ as double integrals (with an integrand which is singular at  $s=s'$) is computationally  cumbersome and, as a result,
the present method is impractical, except for the  case of a circular billiard. In this case 
$A_{ij}(k)$ is a diagonal matrix and its elements can be expressed in terms of products of Bessel and Hankel functions as shown
in Appendix~\ref{sec:circle}.  Equation~(\ref{eq:b8c}) reads then
\begin{equation}
  \text{det}\left[A_{ij}(k)\right] \;\propto\; 
  \prod_{j=-\infty}^{\infty} J_j(k)H_j^{(1)}(k) \;=\;0\;.
  \label{eq:b9}
\end{equation}
The Hankel functions $H_j^{(1)}(k)$ have no real roots and, hence, the energy eigenvalues  for a circular billiard with unit radius are given by the zeros of the integer order Bessel
functions 
\begin{equation}
  J_j\left(k_n\right) \;=\; 0\;,\qquad E_n\; = \; \hbar^2 k_n^2/ 2m\; , 
  \qquad j=0,\pm 1,\pm 2,\ldots
  \label{eq:b10}
\end{equation}
a well known result, which can also be obtained by solving the Schr\"odinger
equation (\ref{eq:i1}) by means of separation of variables\cite{robinett}.
The present derivation of this result demonstrates the equivalence of the
BIE (\ref{eq:b6}) and the stationary Schr\"odinger equation.  Note that
because $J_{-j}(k)=(-1)^jJ_j(k)$ [formula 9.1.5 in Ref.\onlinecite{a&s}] all
the roots corresponding to $j\ne 0$ are doubly degenerate.

\begin{figure}[htbp]
  \centerline{\epsfxsize=4in\epsffile{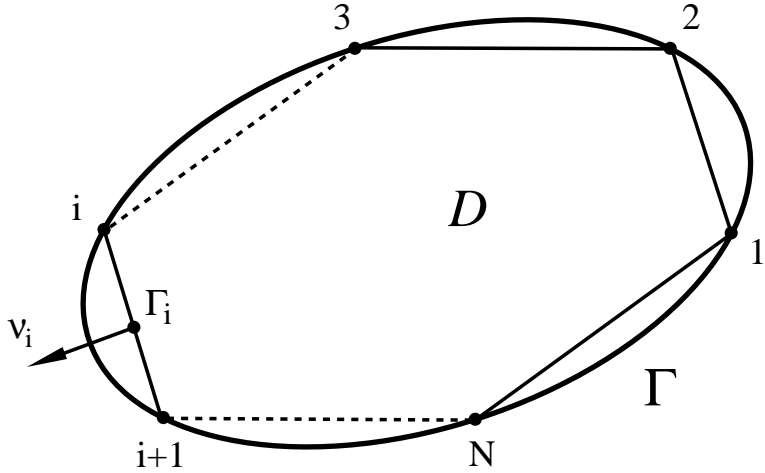}}
  \caption{The billiard boundary \protect$\Gamma$ is approximated by a
    polygon with {\it N} vertices.} 
    \label{fig:discretize}
\end{figure}

\typeout{--- Method II.}
{\bf Method II.}\hspace{2ex} Rather than approximating the BIE in Fourier
space one can approximate it in coordinate space, i.e., one can solve
(\ref{eq:b3}) and not (\ref{eq:b8}). For this purpose one proceeds in two
steps.  First, one approximates the boundary $\Gamma$ by a polygon with $N$
vertices situated on $\Gamma$, as shown in Fig.~\ref{fig:discretize}.
Denoting the segment between vertices $i$ and $i+1$ by $\Gamma_i$ one can
write $\Gamma\approx \cup_{i=1}^N \Gamma_i$, and the BIE can be approximated
by a sum of integrals along the $N$ sides of the polygon. In a second step,
one replaces along each segment $\Gamma_i$ the function
$u(s)\equiv\partial_{\nu}\psi_n(\bbox{r})$ by a constant $u_i$. The BIE is
then replaced by
\begin{equation}
  \label{eq:b11}
  \sum_{i=1}^N u_i \int_{\Gamma_i}\!ds(\bbox{r})\,
  G\left(\bbox{r},\bbox{r'};k\right) \;=\; 0\;.
\end{equation}

Equation~(\ref{eq:b11}) still contains the continuous variable $\bbox{r'}$
which should be eliminated. For this purpose, let us denote the position
vector of the vertex $i$ (see Fig.\ref{fig:discretize}) by $\bbox{s}_i$ and
the position vector of the middle point of $\Gamma_i$ by $\bbox{r}_i=
\left(\bbox{s}_i+\bbox{s}_{i+1}\right)/2$. Then, setting in (\ref{eq:b11})
$\bbox{r'}=\bbox{r}_j$, $j=1,2,\ldots,N$, one arrives at the so-called {\em
  Boundary Element Equation\/} (BEE)\cite{chen-zhou,banerjee}
\begin{equation}
  \label{eq:b12}
  \sum_{i=1}^N u_i \;\Delta{s}_i\!\int_{-\frac{1}{2}}^{\frac{1}{2}}\!d\xi\, 
  G\left(\bbox{r}_i+\xi\Delta\bbox{s}_i,\bbox{r}_j;k\right) \;=\; 0\;,
\end{equation}
where $\Delta\bbox{s}_i\equiv\bbox{s}_{i+1}-\bbox{s}_i$. The above equation
represents a homogeneous system of $N$ linear equations and can be written
\begin{equation}
  \label{eq:b13}
  \sum_{j=1}^N B_{ij}(k) \; u_j \;=\; 0\;.
\end{equation}
The elements of the matrix $B_{ij}(k)$, up to an irrelevant constant factor,
are given by [cf.\ Eq.(\ref{eq:b1})]
\begin{equation}
  \label{eq:b14}
  B_{ij}(k) \;\equiv\; \Delta{s}_j\int_{-\frac{1}{2}}^{\frac{1}{2}}\!d\xi\,
  H_0^{(1)}\left(k\left|\bbox{r}_j-\bbox{r}_i+\xi\Delta\bbox{s}_j\right|\right)  \;.
\end{equation}
In analogy to our previous approach, the (approximate) energy eigenvalues
can be obtained (in terms of $k$) from
\begin{equation}
  \label{eq:b15}
  \text{det}[B_{ij}(k)] \;=\; 0\;, 
\end{equation}
i.e., as the real roots of this equation.

The matrix elements $B_{ij}$ in (\ref{eq:b14}) are expressed as single
integrals in contrast to the matrix elements $A_{ij}$ defined in
(\ref{eq:b8a}) which are expressed in terms of double integrals.  As a
result, Method~II is computationally less demanding than Method~I, but has
nevertheless two unfortunate features. First, the evaluation of the diagonal
matrix elements $B_{ii}$ requires special integration technique due to the
(integrable) singularity of the Green's function at $\xi=0$. Second, in
contrast to Method I where the truncation of the exact, infinite matrix
$A_{ij}$ (defined in the Fourier space) provides us with a natural cutoff
wave vector $K_c$, in case of Method II the relationship between a similar
$K_c$ and the degree of discretization of the boundary (in real space) is
less obvious. 

It should be emphasized that truncation in Fourier space is not quite
equivalent to truncation (discretization of the boundary) in real
space\cite{boasman}. As an empirical rule, if one wishes to calculate energy
eigenvalues corresponding to $k\alt K_c$ accurately, one must take at least
a few (about ten) discretization points for each section of the boundary of
length equal to the corresponding de Broglie wave length $\lambda=2\pi/K_c$.
Thus, the number of discretization points $N$ scales with the length ${\cal
  L}$ of the billiard boundary and the wave vector $K_c$ as follows
\begin{equation}
 N \;\sim\; 10\frac{{\cal L}}{\lambda} \;=\; \frac{10}{2\pi} 
 \left(K_c{\cal L}\right)\;\sim\; K_c{\cal L}. 
  \label{eq:b16}
\end{equation}
Accordingly, accurate calculations of energy eigenvalues corresponding to
sufficiently large $k$ values require a large number of discretization
points $N$ along the boundary $\Gamma$, a condition which leads to large
matrices $B_{ij}$ and, since these matrices are dense, to undesirable
computational efforts.

\typeout{--- Method III.}
{\bf Method III.}\hspace{2ex} The most widely used method for the evaluation
of billiard spectra is based on a non-singular version of the BIE
(\ref{eq:b3}). A simple, but not entirely rigorous\cite{robnik95},
derivation of this method applies the normal derivative operator
$\partial_{\nu'}=\bbox{\nu}(\bbox{r'})\cdot\nabla_{r'}$ to both sides of
Eq.(\ref{eq:bim10}) which, according to definition (\ref{eq:b5}) and with
boundary condition (\ref{eq:b2}) leads to
\begin{equation}
  \label{eq:b17}
  u\left(\bbox{r'}\right) \;=\; - \frac{\hbar^2}{m} \oint_{\Gamma}
  ds(\bbox{r})\, \partial_{\nu'}G\left(\bbox{r},\bbox{r'};E\right)
  u(\bbox{r})\;.
\end{equation}
The integral kernel on the RHS, indeed, is non-singular at
$\bbox{r}=\bbox{r'}$. BIE~(\ref{eq:b17}) is a homogeneous integral equation
with unknown $u(\bbox{r})$; the energy eigenvalues $E$ are given by the
zeros of the corresponding Fredholm determinant [cf.\ Eq.(\ref{eq:b4})],
i.e., as the solutions of
\begin{equation}
  \label{eq:b18}
  \text{det}\left[\delta\left(\bbox{r}-\bbox{r'}\right) + %
  \frac{\hbar^2}{m} %
  \partial_{\nu'}G\left(\bbox{r},\bbox{r'};E\right) \right] \;=\; 0\;.
\end{equation}

Taking into account the explicit form (\ref{eq:b1}) of the free particle
Green's function,  Eq.~(\ref{eq:b17}) can be written [cf.\ Eq.(\ref{eq:ap2-4})]
\begin{equation}
  \label{eq:b17a}
  u\left(\bbox{r'}\right) \;=\; -\frac{ik}{2} \oint_{\Gamma}
  ds(\bbox{r}) \cos\phi\left(\bbox{r'},\bbox{r}\right) 
  H_1^{(1)}\!\left(k|\bbox{r'}-\bbox{r}|\right) u(\bbox{r})\;,
\end{equation}
where
\begin{equation}
  \label{eq:b17b}
  \cos\phi\left(\bbox{r'},\bbox{r}\right) \;\equiv\; %
  \bbox{\nu}(\bbox{r'})\cdot\frac{\bbox{r'}-\bbox{r}}{|\bbox{r'}-\bbox{r}|} 
\end{equation}
is the cosine of the angle between the exterior normal vector to $\Gamma$ at
$\bbox{r'}$ and the unit vector corresponding to $\bbox{r'}-\bbox{r}$. Note
that for $\bbox{r}=\bbox{r'}$ the above scalar product vanishes and,
actually, cancels the singularity due to the Hankel function in the
integrand of the BIE (\ref{eq:b17a}).
 
For a billiard with arbitrary boundary, the above functional determinant
cannot be calculated analytically and one needs to resort to a numerical
solution. For this purpose, one employs the same strategy as in case of
Method II. After discretizing the boundary $\Gamma$, one can replace the BIE
(\ref{eq:b17a}) by the BEE [cf.\ Eq.(\ref{eq:b12})]
\begin{equation}
  \label{eq:b19}
  u_j \;=\; -\frac{ik}{2} \sum_{i=1}^N u_i \Delta s_i
  \int_{-\frac{1}{2}}^{\frac{1}{2}} d\xi\,
   \cos\phi\left(\bbox{r}_j,\bbox{r}_i+\xi\Delta s_i\right) 
   H_1^{(1)}\!\left(k\left|\bbox{r}_j-\bbox{r}_i-%
       \xi\Delta\bbox{s}_i\right|\right)\;,
\end{equation}
where the notations are the same as in the case of Method II. Since the
integrands on the RHS of the above equation are well behaved for all $i,j =
1,2,\ldots,N$, one can approximate the corresponding integrals by the trapezoidal rule. As a result one obtains the  system of
linear equations
\begin{equation}
  \label{eq:b20}
  \sum_{j=1}^N C_{ij}(k)\, u_j \;=\; 0,
\end{equation}
where 
\begin{equation}
  \label{eq:b21}
  C_{ij}(k) \;\equiv\; \delta_{ij}+\frac{ik}{2}\Delta s_j \cos\phi_{ij}
  H_1^{(1)}\!\left(kr_{ij}\right)\;,
\end{equation}
\begin{equation}
  \cos\phi_{ij} \;=\; \bbox{\nu}_i\cdot\frac{\bbox{r}_{ij}}{r_{ij}}\;,
  \qquad
  \bbox{r}_{ij} \;=\; \bbox{r}_i - \bbox{r}_j \;.
  \label{eq:b21a}
\end{equation}
The (approximate) energy eigenvalues can be determined as the roots of the 
determinant of $C_{ij}$
\begin{equation}
  \label{eq:b22}
  \text{det}\left[C_{ij}(k)\right] \;=\; 0\;.
\end{equation}

\typeout{Subsection: Numerical Algorithm for Solving the BEE}
\subsection{Numerical Algorithm for Solving the BEE}
\label{sec:algorithm}

\begin{figure}[htbp]
  \centerline{\epsfxsize=4in\epsffile{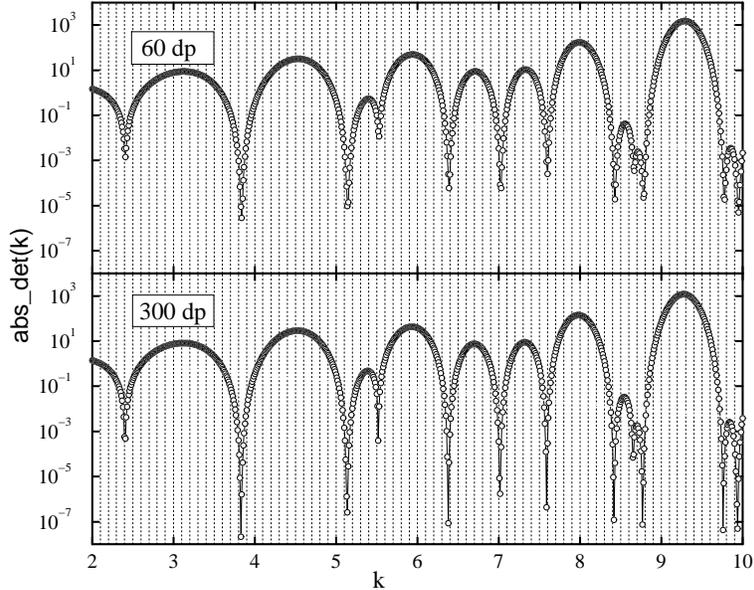}}
  \caption{Plot of {\tt abs\_det}(k) (open circles) for the circle billiard,
    corresponding to 60 and 300 discretization points (dp) of the boundary,
    by using Method III. The positions of the minima approximate the sought
    eigenvalues \protect$k_n$. While the values of the minima depend
    strongly on the degree of discretization of the boundary, the actual
    positions of the minima do not.}
  \label{fig:abs_det}
\end{figure}

Based on the computational methods introduced we have written a {\tt FORTRAN
  77} program which implements the necessary algorithmic steps using the {\em
  SLATEC Common Mathematical Library}\cite{slatec}.  For all three methods
one can employ a common algorithmic framework containing (i) a function {\tt
  det}($k$) which, for an input wave vector $k$, returns the complex value
of the determinant of the corresponding system matrix, i.e., $A_{ij}$,
$B_{ij}$ or $C_{ij}$; (ii) a routine {\tt solve} which calculates
approximately the roots of the equation {\tt det}$(k) = 0$. Once the
function {\tt det}$(k)$ and the corresponding root finder {\tt solve} are
available one can scan the interval of $k$ values of interest (between zero
and the cut-off wave vector $K_c$) to determine the zeros $k_n$ of {\tt
  det}$(k)$ and, hence, the energy eigenvalues $E_n$.  The algorithm may
fail in practice when the separation between two consecutive eigenvalues is
smaller than the scanning step $\Delta k$, i.e., when eigenvalues are nearly
degenerate.  The only way to avoid this error is to use the smallest
affordable $\Delta k$.

\begin{table}
\caption{The first 18 distinct eigenvalues \protect$k_n$ corresponding to
  the circle billiard of unit radius obtained by using Methods I, II
  and III.  The eigenvalues corresponding to Method I represent the
  zeros of the integer Bessel functions \protect$J_j(k)$ and should be
  regarded as {\em exact solutions}. In case of Method II (III) the
  boundary was discretized by employing 60 (300) equally spaced points
  along the circle. The relative error for each approximate solution
  is less than 0.1\%.}
\label{tab:circle}
\begin{tabular}{cccc||ccc}
Method I & Method II & Method III & &
Method I & Method II & Method III\\
\tableline
\tableline
   2.40482&  2.4077 &  2.4053  &  &    8.77148&  8.7800 &  8.7720\\  
   3.83170&  3.8360 &  3.8320  &  &    9.76102&  9.7720 &  9.7615\\  
   5.13562&  5.1415 &  5.1360  &  &    9.93611&  9.9440 &  9.9375\\  
   5.52007&  5.5265 &  5.5206  &  &   10.17347& 10.1855 & 10.1745\\  
   6.38016&  6.3871 &  6.3806  &  &   11.06471& 11.0760 & 11.0655\\  
   7.01558&  7.0233 &  7.0160  &  &   11.08637& 11.0945 & 11.0865 \\ 
   7.58834&  7.5960 &  7.5888  &  &   11.61984& 11.6335 & 11.6200 \\ 
   8.41724&  8.4265 &  8.4175  &  &   11.79153& 11.8055 & 11.7920 \\ 
   8.65372&  8.6640 &  8.6545  &  &   12.22509& 12.2320 & 12.2265 \\     
\end{tabular}
\end{table}

The actual form of {\tt det}$(k)$ depends on the method chosen. In case of
Method I, each matrix element $A_{ij}$ is given by a two-dimensional
integral [see Eq.(\ref{eq:b8a})] with singular and oscillatory integrand
such that the evaluation of {\tt det}$(k)$ would be computationally
extremely demanding and would require special integration routines. Hence,
we did not pursue an implementation of {\tt det}$(k)$ for Method I. In the
case of Methods II and III the function {\tt det}$(k)$ consists of the
following three parts

\begin{itemize}
 
\item[(i)] The subroutine {\tt discretize} which takes as input the data
  necessary to define the actual form of the billiard boundary and the
  number of discretization points $N$ of the billiard boundary; {\tt
    discretize} returns as output the vectors $\bbox{r}_i$, $\bbox{s}_i$ ($i
  = 1,\ldots,N$) [see Fig.~\ref{fig:discretize}] and other useful quantities
  based on them, such as the matrix $\bbox{r}_{ij}=\bbox{r}_i-\bbox{r}_j$,
  the vectors $\Delta\bbox{s}_i = \bbox{s}_{i+1}-\bbox{s}_i$, $\Delta s_i =
  |\Delta\bbox{s}_i|$, $\bbox{\nu}_i =
  \hat{\bbox{z}}\times\Delta\bbox{s}_i/\Delta s_i$ (i.e., the external unit
  vector to the boundary at $\bbox{r}_i$), etc. If one does not want to
  change the degree of discretization of the billiard boundary during the
  successive evaluations of {\tt det}$(k)$, subroutine {\tt discretize}
  should be run only once, namely during the first call of the function {\tt
    det}$(k)$.
  
\item[(ii)] The subroutine {\tt sys\_mat} which evaluates the complex valued
  matrix elements $B_{ij}$ and $C_{ij}$ in case of Method II and III,
  respectively. The $B_{ij}$ are evaluated according to Eq.(\ref{eq:b14})
  employing two {\em SLATEC}\cite{slatec} (more precisely {\em
    QUADPACK}\cite{slatec}) quadrature routines, namely {\tt DQAGS}, for
  calculating the non-diagonal matrix elements, and {\tt DQAWS}, for
  calculating the diagonal matrix elements in which the integrand has a
  logarithmic singularity at $\xi = 0$. The $C_{ij}$ are evaluated according
  to Eqs.(\ref{eq:b21}-\ref{eq:b21a}) in a straightforward way. In both
  cases the Hankel functions can be expressed in terms of the corresponding
  Bessel and Neumann functions for which the double precision {\em SLATEC\/}
  routines {\tt DBESJ0, DBESJ1} and {\tt DBESY0, DBESY1}, respectively, are
  called.
  
\item[(iii)] The function {\tt det}$(k)$ which evaluates the determinant of
  $B_{ij}$ and $C_{ij}$, respectively. For this purpose one employs the {\em
    SLATEC\/} subroutines\cite{slatec} {\tt ZGEFA} (factors a complex matrix
  by using Gaussian elimination) and {\tt ZGEDI} (calculates the determinant
  and the inverse of a complex matrix by using the factors from {\tt ZGEFA}).
\end{itemize}

The function {\tt det}$(k)$ is complex-valued and, therefore, its real roots
$k_n$ (the sought eigenvalues) must be simultaneously zeros of both real and
imaginary parts of this function. Due to the finite discretization of the
boundary, the numerical solutions of the equation {\tt det}$(k)=0$ will be
complex with a (hopefully) small imaginary part. In fact, the magnitude of
the imaginary part of the ``complex eigenvalue'' $\tilde{k}_n$ can be used
to characterize the accuracy of the energy eigenvalues thus determined
through the real part of $\tilde{k}_n$.  To the best of our knowledge, there
exists no public domain subroutine which calculates automatically the roots
of an arbitrary complex function of one complex variable, and as a result
one can make little or no progress at all in the endeavor of constructing a
robust $k_n$ eigenvalue finder algorithm based on the above straightforward
approach. However, there is a relatively simple solution to this problem
which seems to be widely used by practitioners of the
BIM\cite{kitahara,schmit89}. One notes that the zeros of {\tt det}$(k)$ are
also absolute minima for the square of the absolute value of this function,
i.e., of {\tt abs\_det}$(k) \equiv \text{Re} [\text{\tt det}(k)]^2 +
\text{Im}[\text{\tt det}(k)]^2$.  Strictly speaking, the minima should
assume zero values. The discretization of the boundary (or equivalently, the
truncation of the original functional determinant) introduces errors such
that the numerically evaluated minima of {\tt abs\_det}$(k)$, are small, but
not zero; the magnitude of each minimum can be used to distinguish a real
root of {\tt det}$(k)$ from a local minimum of {\tt abs\_det}$(k)$.
Since, numerically, it is much easier to determine the (local) minima of a
real function of a real variable than to determine the roots of a complex
function of a complex variable the suggested approach is much more
convenient for our purpose. Accordingly, {\tt solve} determines actually the
local minima of {\tt abs\_det}$(k)$ by going through a given interval of
wave vectors $k_{min}\le k\le k_{max}$ in steps of $\Delta k$.  Once a
minimum is bracketed, its actual value can be calculated with any desired
accuracy (for a given degree of discretization of the billiard boundary) by
employing, for example, a double precision version of the function {\tt
  brent} from Ref.~\onlinecite{nr-f}.

\typeout{Numerical Results}
\subsection{Numerical Results}
\label{sec:numerics}

As a test of the algorithms described in Sec.~\ref{sec:algorithm} and their
numerical implementation we determine the spectrum of a circular billiard.
In this case Method~I yields the exact energy eigenvalues $k_n$ [cf.\ 
Eq.(\ref{eq:b10})] as the roots of the integer Bessel functions
%
%In this case the energy eigenvalues $k_n$, according to (\ref{eq:b10}) 
%corresponding to solution Method~I, are the roots of the integer Bessel
%function 
%
(these roots are in fact tabulated; see, e.g., Ref.~\onlinecite{a&s}). The
first 18 distinct eigenvalues $k_n$ were also determined by means of
Methods~{II} and {III} described in Sec.~\ref{sec:methods} and are compared
in Table~\ref{tab:circle} with the results of Method~I. In case of Method~II
(III) 60 (300) equally spaced discretization points of the circular boundary
have been employed. The locations of the minima of the function {\tt
  abs\_det}(k) have been determined by scanning the $2\leq k\leq 13$
interval with a step $\Delta k = 0.004$.  Figure~\ref{fig:abs_det}
illustrates the $k$-dependence of {\tt abs\_det}(k), evaluated in the
framework of Method~{III} for two different discretizations of the boundary.
An increase of the number of discretization points from 60 to 300
changes significantly the values, but not the positions of the minima and,
hence, does not affect significantly the values $k_n$.

\begin{figure}[htbp]
  \centerline{\epsfxsize=4in\epsffile{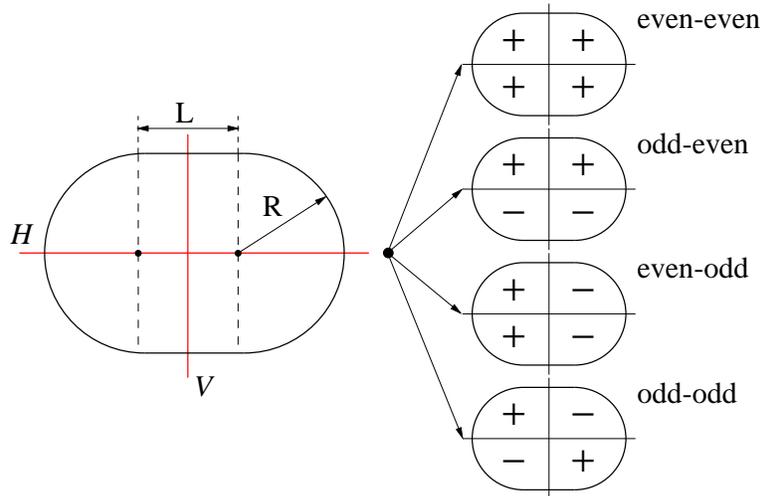}}
  \caption{The stadium billiard is formed by two semicircles of radius
    R connected through two parallel linear segments of length L. The
    two symmetry axes of the stadium are labeled {\it H\/}
    (horizontal) and {\it V\/} (vertical), and the corresponding four
    symmetry classes of the energy eigenfunctions are shown.}
  \label{fig:stadium}
\end{figure}

Table~\ref{tab:circle} demonstrates that the results of both Methods~{II}
and III reproduce the exact eigenvalues to at least three significant digits
for $k\alt K_c$ [cf.\ Eq.(\ref{eq:b16})]. In case of Method~{III}, we have
found that 300 discretization points lead to a precision of better than 1\%
for the 150 lowest eigenvalues of the circular billiard (with unit radius)
corresponding to $k < 35$. For larger $k$ values the density of eigenvalues
increases and, in order to separate adjacent minima of {\tt abs\_det}(k),
one needs to reduce the step size $\Delta k$. The method breaks down for
$k\sim {\cal L}/N$, i.e., when the distance between two consecutive
discretization points of the boundary becomes comparable with the de Broglie
wavelength of the particle, and the only remedy is to increase $N$.

\begin{figure}[htbp]
  \centerline{\epsfxsize=6.4in\epsffile{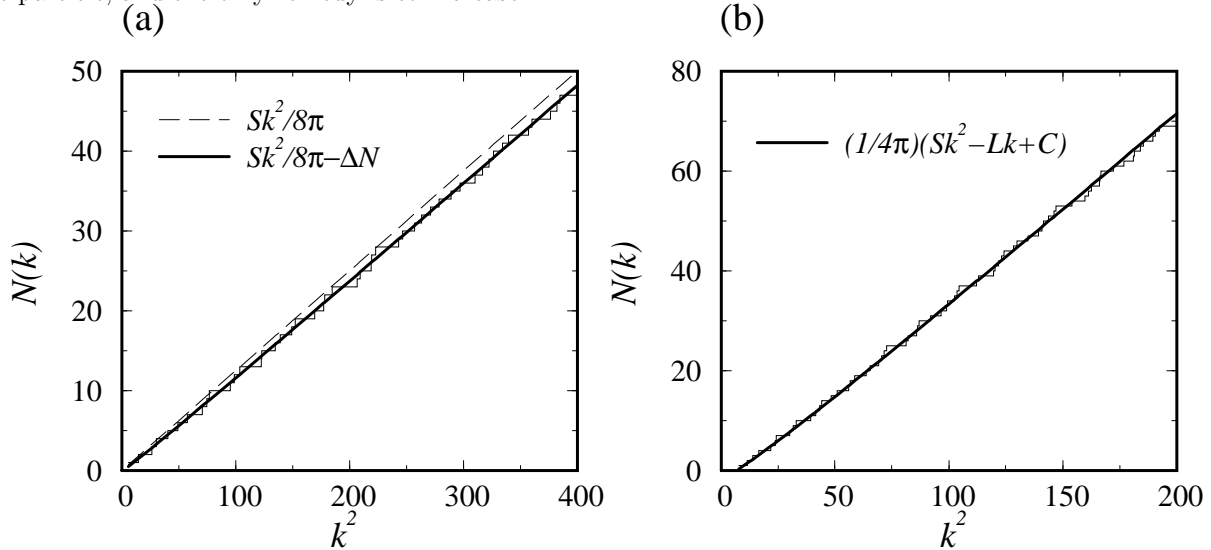}}
  \caption{Spectral staircase \protect$N(k)$ for the lowest 50 (70)
    energy levels of the (a) circle and (b) stadium billiard. In (a) the
    dashed line corresponds to the leading semiclassical {\em Weyl\/} term
    \protect$(Sk^2/4\pi)/2$, where the extra factor of \protect$1/2$
    accounts for the double degeneracy of the energy eigenvalues with
    \protect$m\ne 0$. The solid line is obtained by taking into account the
    perimeter (next to the leading) term in the {\em Weyl\/} formula which
    for the circle billiard is given by \protect$\Delta N = k/4$. In (b) the
    solid line corresponds to the asymptotic {\em Weyl\/} formula with the
    perimeter correction term.}
  \label{fig:stair}
\end{figure}

The program implementing Methods~{II}, III allows one to calculate the
spectra of billiards of arbitrary shapes, for which purpose one needs to
solely alter the coordinates of the discretization points of the billiard
boundary.  As an example, we choose the {\em Bunimovich stadium\/} billiard
depicted in Fig.~\ref{fig:stadium} which consist of two semi-circles (of
radius $R=1$) connected by two parallel linear segments (of length $L$) . We
seek to calculate the lowest few hundred energy eigenvalues of both the
circle and the stadium billiard.

The circle billiard constitutes an {\em integrable\/} system, i.e., the
number of constants of motion (energy and angular momentum) is equal to the
number of degrees of freedom . Its energy eigenstates can be classified
according to symmetry, i.e., by an orbital quantum number $m$, which counts
the nodal lines through the center, and a principal quantum number $n$,
which counts the nodes of the radial wave function, i.e., the nodal
circles\cite{robinett}. In contrast, the stadium billiard, regardless of how
small $L$ is, constitutes a non-integrable, i.e., (strongly) {\em chaotic},
system\cite{berry81,bunimovich}. The study of quantum systems for which the
underlying classical motion is chaotic is a relatively new and still widely
open field of study\cite{lh89,chirikov}. Since it is beyond the scope of the
present article to provide an introduction to quantum chaos, we will content
ourselves with considering without explanation one characteristic which
distinguishes the spectra of non-chaotic (e.g., of a circle billiard) and of
chaotic (e.g., of a stadium billiard) quantum systems, namely the so called
{\em (energy) level spacing distribution\/} $P(s)$. 
By definition\cite{haake,bohigas89}, $P(s) ds$ is the probability that,
given an energy level at $E$, the nearest neighbor energy level is located
in the interval $ds$ about $E+s$. According to {\em Random Matrix
  Theory}\cite{bohigas89,mehta} (RMT), applicable due to a quasi-random
character of the Hamiltonian matrix, quantum systems, as far as the
statistics of their energy spectrum is concerned, in general can be
classified into four universality classes, with well defined and distinct
$P(s)$ level spacing distributions\cite{haake,bohigas89}.  Integrable
systems are described by the Poisson distribution with
\begin{equation}
  \label{eq:num-1}
  P_o(s) \;=\; e^{-s}\;.
\end{equation}
The energy levels of classically chaotic systems, which do not break time
reversal symmetry, (e.g., the stadium billiard) form a Gaussian Orthogonal
Ensemble (GOE) with
\begin{equation}
  \label{eq:num-2}
  P_{GOE}(s) \;=\; \frac{\pi}{2}s\exp\left(-\frac{\pi s^2}{4}\right)\;.
\end{equation}
Further universality classes are the Gaussian Unitary Ensemble (GUE) and the
Gaussian Symplectic Ensemble (GSE); classical chaotic systems which break
time reversal symmetry, e.g., ellipse or stadium billiards in an external
magnetic field, belong to the GUE, while classical chaotic systems which
preserve time reversal symmetry, but break spin rotational symmetry, e.g., a
chaotic billiard in the presence of spin-orbit interaction, belong to the GSE.

\begin{figure}[htbp]
  \centerline{\epsfxsize=6.4in\epsffile{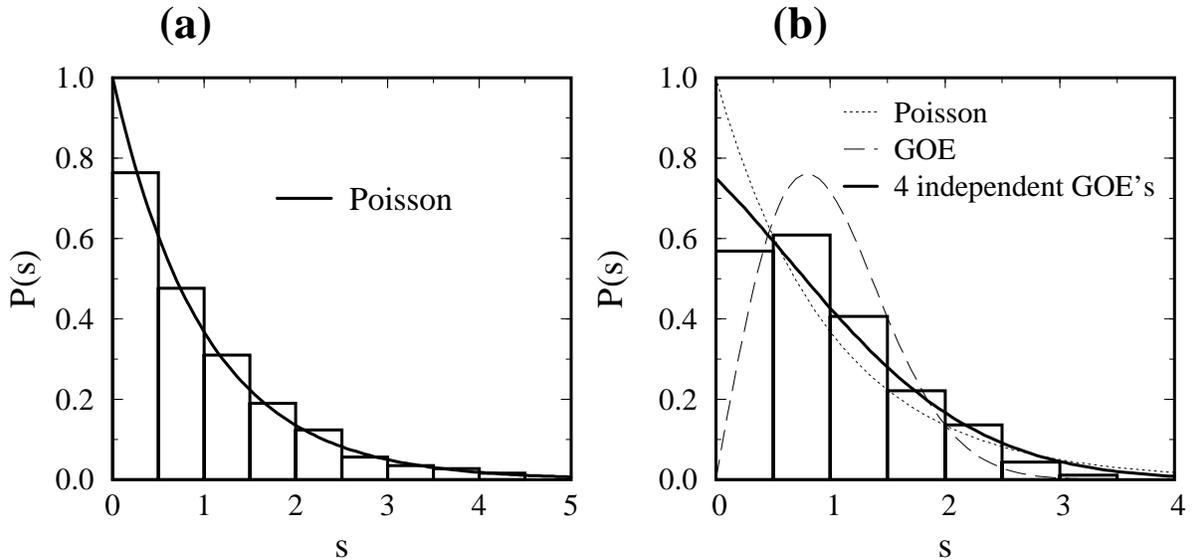}}
  \caption{Histogram of the energy level spacing distribution \protect$P(s)$
    for the (a) circle and (b) stadium billiards. In (a) the histogram is
    constructed from the lowest 1,200 energy levels of the circle billiard.
    The solid line corresponds to the Poisson prediction for the level
    spacing distribution. In (b) the lowest 600 energy levels have been used
    to construct the histogram. The dotted (dashed) line represents the
    Poisson (GOE) prediction for \protect$P(s)$. The histogram is best
    approximated by the superposition of 4 independent GOE distributions
    (solid line) which correspond to the same number of distinct symmetry
    classes of the energy eigenstates in a stadium billiard.}
  \label{fig:ps}
\end{figure}

\begin{figure}[htbp]
  \centerline{\epsfxsize=6.4in\epsffile{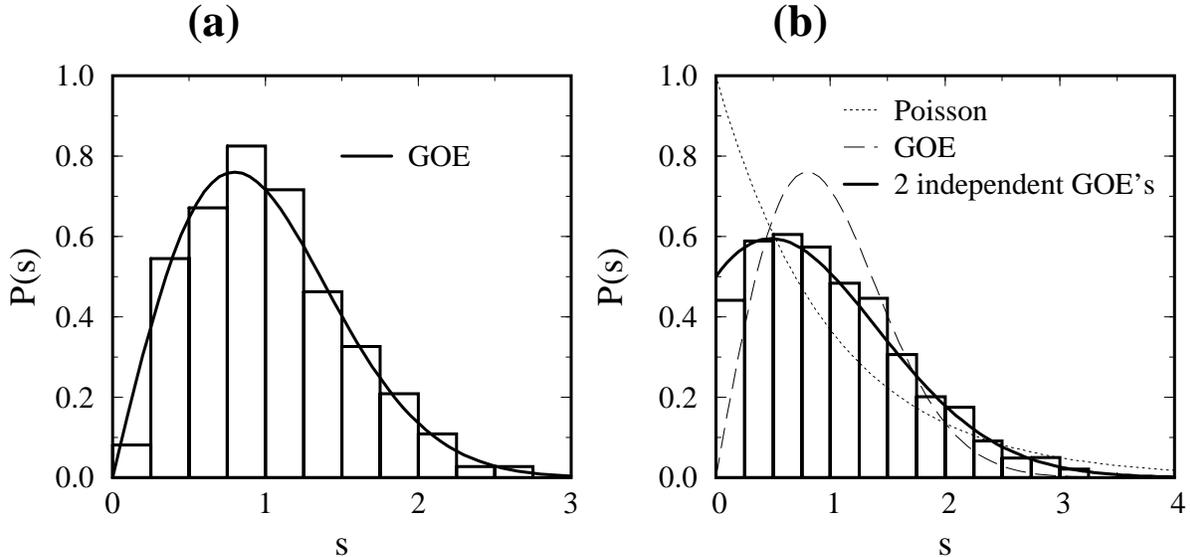}}
  \caption{Histogram of the energy level spacing distribution 
    \protect$P(s)$ for the (a) quarter- and (b) half-stadium
    billiards. The level spacing distribution for a quarter-
    (half-) stadium is well approximated by the GOE (two independent
    sequences of GOE) distribution function.}
  \label{fig:ps-}
\end{figure}

Poisson and GOE distributions are distinguished most clearly near $s=0$,
since $P_o(0)=1$ constitutes the maximum of $P_o$ while $P_{GOE}(0)=0$
constitutes the minimum of $P_{GOE}$; neighboring energy levels are
likely to attract (repel) each other in the case of integrable (chaotic)
systems. We want to show that the level spacing distribution evaluated by
means of the BIM for circle and stadium billiards satisfies the Poisson and
GOE distribution, respectively.  For this purpose one needs to calculate at
least a few hundred of the lowest energy levels without actually missing any
energy levels since such misses would distort the energy level spacing
distribution.  The quality of the calculations, in particular in the case of
the stadium billiard, can be judged from a comparison of the obtained {\em
  (energy) staircase function\/} ${\cal N}(E)$ (which gives the number of
quantum states with energy less or equal to $E$) with the corresponding {\em
  Weyl}-type formula\cite{bohigas89,gutzwiller}
\begin{equation}
  \label{eq:num-3}
  \langle{\cal N}(E)\rangle \;=\; \frac{1}{4\pi}%
  \left(S\,E - {\cal L}\,\sqrt{E} + C\right)\;,
\end{equation}
where $S$ and  ${\cal L}$ are the area and perimeter of the billiard, and
$C$ is a constant related to the geometry and topology of the billiard
boundary. Presently, we employ units in which $\hbar^2/2m$
is equal to one; thus, e.g., $E=k^2$. Also, in the numerical results
reported below we have chosen $L=R=1$ (see Fig.~\ref{fig:stadium}).

Strictly speaking Eq.(\ref{eq:num-3}) is valid only in the semi-classical
($E\rightarrow\infty$) limit, but in practice it turns out that one can
apply Weyl's formula even at the lower end of the energy spectrum. Our
results for the staircase function ${\cal N}(k)$, corresponding to the first
50 (70) distinct energy levels of the circle (stadium) billiard, are
presented in Fig.~\ref{fig:stair}. In the case of the circle billiard a
complication arises due to the fact that all the energy levels with angular
momentum $m\ne 0$ are doubly degenerate.  A simple remedy to this problem is
to assume that the fraction of energy levels corresponding to $m=0$ is
negligible in comparison to those with $m\ne 0$, and that the double
degeneracy can be accounted for by dividing the RHS of Eq.(\ref{eq:num-3})
by two.

Based on the good agreement between ${\cal N}(E)$ and $\langle{\cal
  N}(E)\rangle$ shown in Fig.~\ref{fig:stair}, one may conclude that all
energy levels have been accounted for. A similar analysis for the first 600
energy levels showed that at most a few percent of the energy levels might
been missed. This conclusion is independent of the method chosen, i.e., of
Methods~{II} and III.

\begin{table}
\caption{Wave vector eigenvalues \protect$k_n\;(< 10)$ for (a)
quarter-, (b) horizontal half-, (c) vertical half- and (d) full stadium
billiards. The quarter stadium has only odd-odd energy eigenstates, the
horizontal (vertical) half stadium has both odd-odd and odd-even (even-odd)
eigenstates, while the full stadium has eigenstates which belong to
all four symmetry classes. The symmetry of each eigenstate of the
(full) stadium billiard can be identified based on this table as
explained in the text. The dash in each column indicates that the
corresponding eigenvalue is absent for that system.}
\label{tab:stadium}
\begin{tabular}{ccccc||ccccc}
quarter & horizontal half & vertical half & full    &  &  quarter &
horizontal half & vertical half & full \\
stadium & stadium    & stadium    & stadium &  &  stadium & stadium & stadium & stadium\\ 
\tableline
\tableline
%---------------------------------------------------------------------------------------
   --   &    --    &  2.7784  &  2.7785  &  &  7.5231  &  7.5238  &  7.5238  &  7.5240 \\
\tableline 
  --    &  3.4037  &    --    &  3.4037  &  &    --    &  7.6642  &    --    &  7.6640 \\ 
\tableline                         
  --    &    --    &    --    &  3.7211  &  &    --    &    --    &    --    &  7.9760 \\ 
\tableline                                 
4.0564  &  4.0565  &  4.0565  &  4.0566  &  &    --    &    --    &    --    &  8.0945 \\ 
\tableline                                 
  --    &    --    &  4.6786  &  4.6785  &  &    --    &    --    &  8.3192  &  8.3200 \\ 
\tableline                                 
  --    &  4.8800  &    --    &  4.8800  &  &    --    &  8.3989  &    --    &  8.3985 \\ 
\tableline                                 
  --    &    --    &    --    &  4.9223  &  &  8.4639  &  8.4642  &  8.4639  &  8.4640 \\ 
\tableline                                 
  --    &    --    &  5.4931  &  5.4935  &  &    --    &    --    &  8.5200  &  8.5200 \\ 
\tableline                                 
  --    &    --    &    --    &  5.6360  &  &    --    &    --    &  9.0100  &  9.0105 \\
\tableline                                 
5.7456  &  5.7456  &  5.7456  &  5.7455  &  &    --    &    --    &    --    &  9.0600 \\ 
\tableline                                 
  --    &    --    &    --    &  6.2714  &  &  9.2641  &  9.2650  &  9.2650  &  9.2655 \\ 
\tableline                                 
  --    &  6.4387  &    --    &  6.4385  &  &    --    &  9.2890  &    --    &  9.2895 \\ 
\tableline                                 
  --    &    --    &  6.5743  &  6.5751  &  &    --    &    --    &    --    &  9.3200 \\ 
\tableline                                 
  --    &  6.6493  &    --    &  6.6495  &  &    --    &  9.5900  &    --    &  9.5895 \\ 
\tableline                                 
6.9526  &  6.9531  &  6.9526  &  6.9528  &  &    --    &    --    &  9.8281  &  9.8280 \\ 
\tableline                                 
  --    &    --    &  7.1350  &  7.1352  &  &  9.9481  &  9.9481  &  9.9481  &  9.9480 \\ 
\tableline                                 
  --    &    --    &    --    &  7.4815  &  &    --    &    --    &    --    &  9.9720 \\ 
%---------------------------------------------------------------------------------------
\end{tabular}
\end{table}

For a proper analysis of the energy level statistics we linearly scale the
set of energy eigenvalues such that for the resulting sequence the mean level
spacing is uniform and equal to unity. This transformation, known as
``unfolding the spectrum''\cite{haake,bohigas89}, is commonly achieved by
replacing the original set of eigenenergies $E_n = k_n^2$ by
\begin{equation}
  \label{eq:num-5}
  \tilde{E}_n \;=\; \langle{\cal N}(E_n)\rangle\;.
\end{equation}
The unfolded spectrum now can be used to calculate the nearest level
spacings $s_n = \tilde{E}_{n+1}-\tilde{E}_n$, which fluctuate about their
mean value equal to one. Finally, a normalized histogram of the $s_n$ series
gives a rough representation of the distribution function $P(s)$.  The
resulting distributions $P(s)$ for the circle and stadium billiards are
shown in Fig.~\ref{fig:ps}. In the case of the circle billiard the obtained
histogram agrees very well with the expected Poisson distribution
Eq.(\ref{eq:num-1}).  However, in the case of the stadium billiard the
$P(s)$ histogram does not resemble a GOE distribution, in particular, the
distribution exhibits a clear absence of level repulsion, i.e., $P(s)$ does
not vanish for $s\rightarrow 0$.

The deviation of $P(s)$ from a GOE distribution arises due to the fact that
the stadium billiard, even though it is chaotic, exhibits a geometrical
symmetry with two symmetry planes\cite{bohigas84}, as shown in
Fig.~\ref{fig:stadium}.  Accordingly, the stationary states fall into four
distinct symmetry classes, according to their parity (i.e., either {\em
  odd\/} or {\em even\/}) with respect to reflection at the two planes. As a
result, the stadium billiard spectrum is composed of four independent family
of states, each of which is expected to conform to a GOE distribution. A
general expression for the level spacing distribution function $P^{(N)}(s)$
corresponding to the superposition of $N$ independent spectra with GOE
statistics is derived in Appendix~\ref{sec:sup-GOE}. Thus, the level spacing
distribution corresponding to the full stadium billiard is given by
Eq.(\ref{eq:sup-GOE-6}) with $N=4$, i.e.,
\begin{eqnarray}
\label{eq:num5a}
 P^{(4)}(s) &\;=\;& \frac{\partial^2}{\partial s^2} 
 \left[\text{erfc}\left(\frac{\sqrt{\pi}s}{8}\right)\right]^4 \\
 &\;=\;&
 \frac{3}{4}\,\exp\left(-\frac{\pi\,s^2}{32}\right)\,
 \left[\text{erfc}\left(\frac{\sqrt{\pi}\,s}{8}\right)\right]^2 +
 \frac{\pi}{32}\,\exp\left(-\frac{\pi\,s^2}{64}\right)\,
 \left[\text{erfc}\left(\frac{\sqrt{\pi}\,s}{8}\right)\right]^3 \;,\nonumber
\end{eqnarray}
where $\text{erfc}(z)$ is the complementary error function\cite{a&s}.
Comparison of the numerically determined $P(s)$ with the distribution
(\ref{eq:num5a}) in Fig.~\ref{fig:stadium} is indeed satisfactory.  The
small values of $P(s)$ for small $s$-values, i.e., values below the
prediction by (\ref{eq:num5a}), are likely due to an omission of ``nearly
  degenerate'' eigenvalues by our spectrum finder routine (see also below).

In order to check the assertion made about the symmetry classes of the
energy eigenstates, and about the corresponding level spacing distributions,
 we have calculated and analyzed also the energy
spectrum of a quarter stadium, and of the upper (horizontal) half and right
(vertical) half stadium billiards, as well.  The results are shown in
Fig.~\ref{fig:ps-}. Indeed, the $P(s)$ histogram for the quarter stadium,
which accommodates all the eigenstates with odd--odd symmetry (see
Fig.~\ref{fig:stadium}) conforms to a GOE distribution. On the other hand,
for each of the two half stadiums, with eigenstates which belong to two
distinct symmetry classes, namely odd--odd and odd--even (even--odd) in the
case of horizontal (vertical) half stadiums, the level spacing distribution
histogram is in good agreement with the theoretical prediction of the
superposition of two independent GOE's as described by
Eq.(\ref{eq:sup-GOE-6}) with $N=2$, i.e.,
\begin{eqnarray}
\label{eq:num5b}
 P_2(s) &\;=\;& \frac{\partial^2}{\partial s^2} 
 \left[\text{erfc}\left(\frac{\sqrt{\pi}s}{4}\right)\right]^2 \\
 &\;=\;&
 \frac{1}{2}\,      \exp\left(-\frac{\pi\,s^2}{8}\right) +
 \frac{\pi\,s}{8}\, \exp\left(-\frac{\pi\,s^2}{16}\right)\, 
 \text{erfc}\left(\frac{\sqrt{\pi}\,s}{4}\right)\;. \nonumber
\end{eqnarray}

It should be noted that one can also identify the symmetry of each energy
level of the stadium billiard. For this purpose one needs the
energy spectrum of the full-, quarter-, horizontal half- and vertical half
stadiums. These eigenenergies, corresponding to $k_n < 10$, are listed in a
convenient way in Table~\ref{tab:stadium}. The odd--odd eigenvalues can be
simply read out from the column which contains the spectrum of the quarter
billiard. Obviously, this eigenvalues belongs also to the other three
billiards under consideration. The odd--even (even--odd) eigenvalues can be
obtained from the spectrum of the horizontal (vertical) half stadium by
removing from the corresponding spectrum all the already known odd--odd
eigenvalues. Finally, all the energy levels of the full stadium which have
not been accounted for so far have even--even parity.

We conclude this section with a few comments on the distribution function
[Eq.(\ref{eq:sup-GOE-6})] 

\[ 
  P^{(N)}(s) \;=\; \frac{\partial^2}{\partial\,s^2}
  \left[\text{erfc}\left(\frac{\sqrt{\pi}}{2}\frac{s}{N}\right)\right]^N 
\]
describing the superposition of $N$ GOE distributions.  
For $N=1$ one recovers the GOE distribution function (\ref{eq:num-2})
wich is normalized and yields a mean level spacing equal to one. In
the limit $N\rightarrow\infty$, by using the the series expansion\cite{a&s}
$\text{erfc}(z) = 1-(2/\sqrt{\pi})\,z + {\cal O}\left(x^3\right)$ and the
definition\cite{a&s} $\exp(-x) = \lim_{N\rightarrow\infty}(1-x/N)^N$, one
arrives at $P^{(\infty)}(s) = \exp{(-s)}$, which is exactly the Poisson
distribution (\ref{eq:num-1}). This result is a particular case of the
theorem according to which the level spacing distribution of the
superposition of infinitely many independent spectra (with {\em arbitrary\/}
level spacing distributions) is always Poisson like\cite{haake,bohigas89}.

\begin{figure}[htbp]
  \centerline{\epsfxsize=6in\epsffile{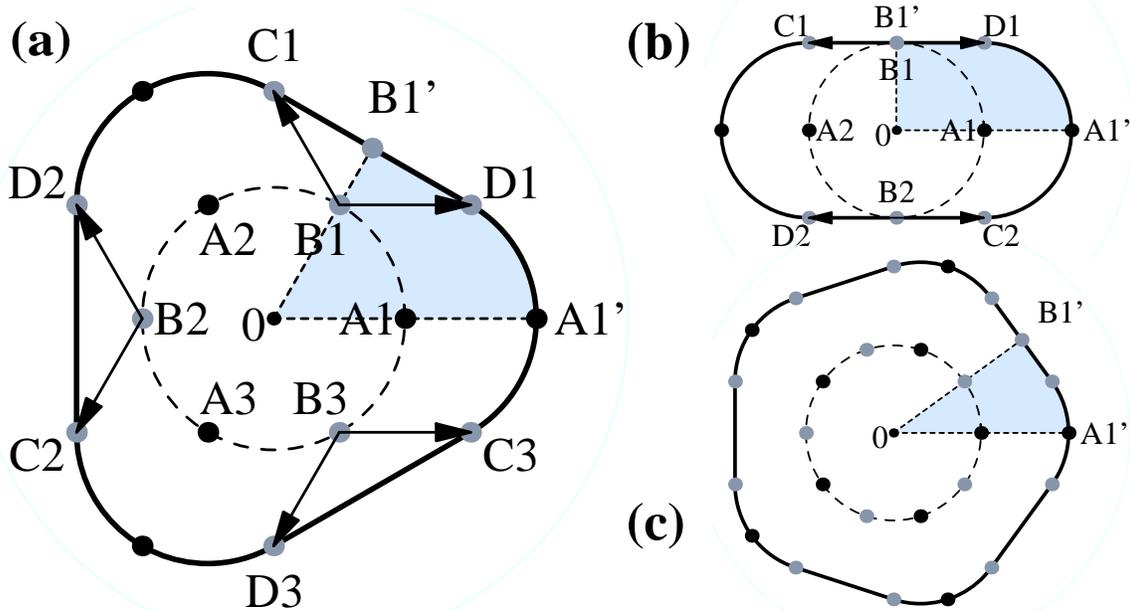}}
  \caption{(a) Geometrical construction of the deformed billiard ($N=3$),
    starting form the unit circle billiard. (b) The stadium billiard as an
    $N=2$ deformed billiard. (c) Deformed billiard for $N=5$. The
    highlighted regions correspond to ``elementary sectors'' for which
    simple GOE level spacing distribution is expected.}
  \label{fig:circle_deform}
\end{figure}

Inspired by the billiard stadium problem, we propose a closely related
numerical experiment which tests the appearance of the distribution
$P^{(N)}(s)$ (\ref{eq:sup-GOE-6}).  For this purpose we determine the energy
levels corresponding to a deformation of the circle billiard involving an
$N$-fold symmetry axis.  Let us consider $N$ equidistant points $A_i$,
$i=1,...,N$ on the unit circle, with center $O$.  $B_i$ is the midpoint of
the arc of circle $A_iA_{i+1}$.  We construct then points $C_i$ and $D_i$ by
translating $B_i$ with the vectors $\epsilon\cdot\bbox{OA_{i+1}}$ and
$\epsilon\cdot\bbox{OA_i}$, respectively.  The parameter $\epsilon$ controls
the degree of the deformation. The deformed billiard is defined by the
linear segments $D_iC_i$ and the arcs of circle $C_iD_{i+1}$ with unit
radii. The new billiard, for $N=3$ and $\epsilon =1$, is illustrated in
Fig.~\ref{fig:circle_deform}a. In the limit $\epsilon\rightarrow 0$ one
recovers the original circle billiard.  For $N=2$, the new billiard is
actually the stadium billiard, as shown in Fig.~\ref{fig:circle_deform}b.
For $N>2$, the billiards can be regarded as a generalization of the stadium
billiard; this is illustrated for another case, $N=5$, in
Fig.~\ref{fig:circle_deform}c.

\begin{figure}[htbp]
  \epsfxsize=6.4in\epsffile{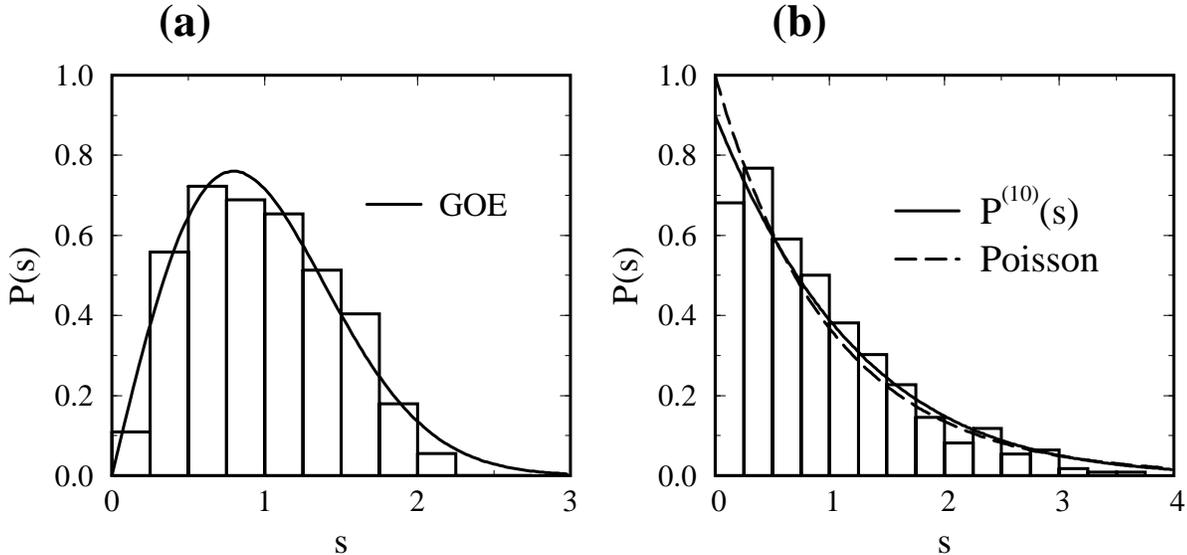}
  \caption{ Histogram of the energy level spacing distribution 
    \protect$P(s)$ corresponding to \protect$N=5$ for (a) an elementary
    sector, and (b) full deformed billiards. The level spacing distribution
    for the elementary sector (full) deformed billiard is well approximated
    by the GOE \protect$\left[P^{(10)}(s)\right]$ distribution function
    (solid line).  \protect$P(s)$ for the full billiard differs only
    slightly from the Poisson distribution (dashed line).}
  \label{fig:n=5}
\end{figure}

For a given $N$, the deformed billiard possesses $N$ symmetry planes and,
therefore, the corresponding stationary states fall into $2N$ distinct
symmetry classes, according to their parity with respect to reflection at
these $N$ planes. Proceeding as in the case of the stadium billiard, one can
divide the deformed billiard into $2N$ elementary sectors (see the highlighted
regions in Fig.~\ref{fig:circle_deform}). The energy levels of a single
sector should have an energy spectrum with GOE statistics. This is, indeed,
born out of a BEM calculation as shown by the corresponding match with a GOE
distribution in Fig.~\ref{fig:n=5}a in case of a single $N=5$ sector.  A
billiard formed by attaching $n$ ($n\le 2N$) such elementary sectors should
exhibit a level spacing distribution given by $P^{(n)}(s)$, while the level
spacing distribution corresponding to the full deformed billiard should
conform to $P^{(2N)}(s)$. The level spacing distribution of an $N=5$
billiard conforms well to the distribution $P^{(10)}(s)$ as seen in
Fig.~\ref{fig:n=5}b.

In the limit $N\rightarrow\infty$ the deformed billiard becomes a circle of
radius $1+\epsilon$ as one can infer readily from the construction presented
in Fig.~\ref{fig:circle_deform}.  The suggested billiards produce then level
spacing distributions which, due to
$\lim_{N\rightarrow\infty}P^{(N)}(s)=P_o(s)$, conform to a Poisson
distribution.  This is to be expected, of course, since this distribution
governs the spectrum of a circle billiard.  One can recognize in
Fig.~\ref{fig:n=5}b that already in the case $N\, = \, 5$ the level spacing
distribution resembles the Poisson distribution.

Many further billiards can be constructed in a similar way. For the case of
classical systems, a family of billiards which exhibit chaotic as well as
mixed chaotic and regular motion have been studied in
Ref.~\onlinecite{drw96}. The application of the BEM to determine level
statistics as well as wave functions for the mixed system might reveal some
surprising behavior.
%---------------------------------------------------------------

\typeout{Other Examples}
\section{Other Examples}
\label{sec:examples}

In this section we wish to present two other examples in which the BIM can
be applied. Both examples exhibit the features mentioned at the end of
Sec.~\ref{sec:bim}: (i) the corresponding Green's function is known
analytically; (ii) the boundary condition at $\Gamma$ assumes a simple form.
Due to lack of space we shall only outline the BIM treatment of these
examples. The interested reader is encouraged to work out further details,
including the statistical analysis of the obtained data, in analogy to the
quantum billiard case presented in the previous section.

%___________________________
\typeout{Finite Potential Well}
\subsection{Finite Potential Well}
\label{sec:pot-well}

As a first example let us consider a particle trapped inside a
two-dimensional potential well defined by a finite potential increase at the
boundary, described by the potential

\begin{equation}
  \label{eq:pot-well-1}
  V(\bbox{r}) \;=\; 
  \left\{
    \begin{array}{llll}
      0   \hspace*{1ex}&,\hspace{1.5ex}&\text{for}\hspace{2ex}& \bbox{r}\in
      {\cal D}_i \\ 
      V_o  &, &\text{for}& \bbox{r}\in {\cal D}_o \;.
    \end{array}
  \right.  
\end{equation}
Here ${\cal D}_{i/o}$ represents the inner/outer region determined by a
closed boundary $\Gamma$ of arbitrary shape. The depth of the potential well is
$V_o\,(>0)$. The energy spectrum for this system has a discrete part for
$E_n < V_o$, and a continuous part for $E>V_o$. The quantum billiard
studied in the previous section can be regarded as a limiting case of the
present case corresponding to $V_o\rightarrow\infty$.

For the purpose of calculating the discrete energy eigenvalues of the system
one applies the BIM presented in Sec.~\ref{sec:bim} for both inner (${\cal
  D}_i$) and outer (${\cal D}_o$) regions. As a result one obtains a set of
two coupled BIE's; the two unknown functions are the wave function $\psi_n$
and its outward (with respect to the inner region ${\cal D}_i$) normal
derivative $\partial_{\nu}\psi_n$ along $\Gamma$.

For ${\cal D}\equiv{\cal D}_i$, in analogy to Eq.(\ref{eq:bim10}), the
corresponding BIE reads

\begin{mathletters}
\label{eq:pot-well-2}
\begin{equation}
  \label{eq:pot-well-2a}
  \psi^{(i)}_n(\bbox{r'}) \;=\; \frac{\hbar^2}{m} {\cal P}\oint_{\Gamma}\!
 ds(\bbox{r}) \left[\psi^{(i)}_n(\bbox{r})\partial_{\nu}% 
 G^{(i)}\left(\bbox{r},\bbox{r'};E_n\right)-
 G^{(i)}\left(\bbox{r},\bbox{r'};E_n\right)%
 \partial_{\nu}\psi^{(i)}_n(\bbox{r})\right]\;,
\end{equation}
with the Green's function 
\begin{equation}
  \label{eq:pot-well-2b}
    G^{(i)}\left(\bbox{r},\bbox{r'};E_n\right) \;= \; -\frac{im}{2\hbar^2}
    H_0^{(1)}\left(k_n\left|\bbox{r}-\bbox{r'}\right|\right) \;,
    \qquad
    k_n\;=\; \sqrt{2mE_n}/\hbar\;.
\end{equation}
\end{mathletters}

The ``exterior problem'' ${\cal D}\equiv{\cal D}_o$ requires a more careful
treatment due to the fact that ${\cal D}_o$ is unbounded. One can circumvent
this difficulty by considering instead a finite region ${\cal D}_{\rho}$
delimited by $\Gamma$ inside and by a circle ${\cal C}_{\rho}$ with a very
large radius $\rho$ outside, the center of the later located somewhere
inside the region ${\cal D}_i$; evidently,
$\lim_{\rho\rightarrow\infty}{\cal D}_{\rho} = {\cal D}_o$. Thus, when we
apply Green's formula to obtain the BIE an extra term results in
(\ref{eq:bim10}) due to the circle ${\cal C}_{\rho}$. However, this
extra term vanishes in the limit $\rho\rightarrow\infty$ because for bound
states (the only ones we are interested in) both the wave function
$\psi^{(o)}_n$ and its gradient $\nabla\psi^{(o)}_n$ vanish exponentially at
infinity. Hence, the corresponding BIE becomes

\begin{mathletters}
\label{eq:pot-well-3}
\begin{equation}
  \label{eq:pot-well-3a}
  \psi^{(o)}_n(\bbox{r'}) \;=\; - \frac{\hbar^2}{m}\,{\cal P}\!\oint_{\Gamma}\!
  ds(\bbox{r}) \left[\psi^{(o)}_n(\bbox{r})\partial_{\nu}
  G^{(o)}\left(\bbox{r},\bbox{r'};E_n\right)-
  G^{(o)}\left(\bbox{r},\bbox{r'};E_n\right)%
  \partial_{\nu}\psi^{(o)}_n(\bbox{r})\right]\;,
\end{equation}
with the Green's function (which is finite for $|\bbox{r}-\bbox{r'}|
\rightarrow\infty$)
  \begin{equation}
    \label{eq:pot-well-3b}
    G^{(o)}\left(\bbox{r},\bbox{r'};E_n\right) \;= \; -\frac{i\,m}{2\hbar^2}
    H_0^{(1)}\left(iq_n\left|\bbox{r}-\bbox{r'}\right|\right) \;=\;
    -\frac{1}{\pi} K_0\left(q_n\left|\bbox{r}-\bbox{r'}\right|\right)\;,
  \end{equation}
\[  
    q_n\;=\; \sqrt{2m\left(V_o-E_n\right)}/\hbar\;.
\]  
\end{mathletters}%
Here $K_0(z)$ is a Bessel function of imaginary argument\cite{a&s} (see also
Appendix~\ref{sec:free-particle}).
Note the minus sign on the RHS of Eq.(\ref{eq:pot-well-3a}) which accounts
for the opposite orientation of the exterior normal unit vectors
corresponding to ${\cal D}_i$ and ${\cal D}_o$.

Since the wave function and its normal derivative must be continuous across
$\Gamma$, i.e.,

\begin{mathletters}
\label{eq:pot-well-4}
\begin{equation}
  \label{eq:pot-well-4a}
  \psi^{(i)}_n(\bbox{r}) \;=\; \psi^{(o)}_n(\bbox{r}) \;\equiv\;
  \psi_n(\bbox{r}) \;,
\end{equation}
\begin{equation}
  \label{eq:pot-well-4b}
  \partial_{\nu}\psi^{(i)}_n(\bbox{r}) \;=\; 
  \partial_{\nu}\psi^{(o)}_n(\bbox{r}) \;\equiv\;
  \partial_{\nu}\psi_n(\bbox{r}) \;,
\end{equation}
\end{mathletters}%
Eqs.(\ref{eq:pot-well-2}-\ref{eq:pot-well-3}) form a set of coupled BIE's
with respect to the unknown functions $\psi_n$ and $\partial_{\nu}\psi_n$.

The numerical calculation of the energy levels of a particle in a finite
two-dimensional potential well proceeds similarly as in the case of a
quantum billiard. The steps to be filled in are the same as those discussed
in Secs.~\ref{sec:methods},\ref{sec:algorithm}. Note, however, that due to
the simultaneous presence of both $\psi_n$ and $\partial_{\nu}\psi_n$ in the
BIE's, only Method~{II} can be applied in this particular case.

%--------------------------------------------------
\typeout{Quantum Billiard in a Magnetic Field}
\subsection{Quantum Billiard in a Magnetic Field}
\label{sec:mag-field}

As a second example, let us consider a charged particle confined to a
two-dimensional billiard with $V_o\rightarrow\infty$ [cf.\ 
Eq.(\ref{eq:pot-well-1})], in the presence of a uniform magnetic field
$\bbox{B}$ perpendicular to the plane of motion. The Hamiltonian for such
{\em quantum billiard in a magnetic field} is given by [cf.\ 
Eq.(\ref{eq:bim1})]
\begin{equation}
  \label{eq:mf-1}
  \hat{H} \;=\; \frac{1}{2m}\left(\bbox{p}-q\bbox{A}\right)^2 +
  V(\bbox{r})\;,
\end{equation}
where $\bbox{p}=-i\hbar\nabla$ is the momentum operator, $q$ is the electric
charge of the particle, $\bbox{A}(\bbox{r})$ is the vector potential
($\bbox{B}=\nabla\times\bbox{A}$) and $V(\bbox{r})$ is the scalar potential
as given by Eq.(\ref{eq:pot-well-1}). The energy spectrum of the system can
be determined by solving the Schr\"odinger equation (\ref{eq:i1}) subject to
the Dirichlet boundary condition (\ref{eq:b2}). To derive the corresponding
BIE we rewrite the Hamiltonian (\ref{eq:mf-1}), recalling that for a
static magnetic field $\nabla\cdot\bbox{A}=0$,
\begin{equation}
  \label{eq:mf-1b}
  \hat{H} \;=\; -\frac{\hbar^2}{2m}\nabla^2 + \frac{q^2A^2}{2m} -
  i\,\frac{q\hbar}{m} \bbox{A}\cdot\nabla \;,
\end{equation}
and define the Green's function $G(\bbox{r},\bbox{r'};E)$ as the solution of 
\begin{equation}
  \label{eq:mf-2}
  \left[E-\hat{H}^*(\bbox{r})\right] G\left(\bbox{r},\bbox{r'};E\right) \;=\;
  \delta\left(\bbox{r}-\bbox{r'}\right)\;,
\end{equation}
where $\hat{H}^*$ is the complex conjugate of the Hamiltonian
(\ref{eq:mf-1b}). Note that $\hat{H}^*\ne\hat{H}$, which implies that the
magnetic field breaks time reversal symmetry\cite{haake}. 

Using the same strategy as in Sec.~\ref{sec:bim}, one can derive the
following BIE
\begin{eqnarray}
  \label{eq:mf-3}
  \psi_n(\bbox{r'}) &\;=\;& \frac{\hbar^2}{m}
  {\cal P}\oint_{\Gamma}\!ds(\bbox{r}) 
  \left[\psi_n(\bbox{r})\partial_{\nu}G\left(\bbox{r},\bbox{r'};E_n\right) - %
    G\left(\bbox{r},\bbox{r'};E_n\right)%
    \partial_{\nu}\psi_n\left(\bbox{r}\right)\right]\nonumber\\
  &\;+\;& i\,\frac{2q\hbar}{m}\oint_{\Gamma}\!ds(\bbox{r}) A_{\nu}(\bbox{r})
  G\left(\bbox{r},\bbox{r'};E_n\right) \psi_n(\bbox{r})\;,
\end{eqnarray}%
where $A_{\nu}\equiv\bbox{\nu}\cdot\bbox{A}$. Since the wave function
$\psi_n$ vanishes along the boundary of the billiard $\Gamma$ [cf.\ 
Eq.(\ref{eq:b2})] the last term in Eq.(\ref{eq:mf-3}) can be dropped. As a
result, we obtain formally the same BIE as in the field-free case, namely
Eq.(\ref{eq:b3}), or equivalently Eq.(\ref{eq:b18}).  Hence, the energy
levels of a quantum billiard in a magnetic field can be determined as
described in Sec.~\ref{sec:billiard}. The only difference is that, instead
of the free particle Green's function, the Green's function of a charged
particle in magnetic field needs to be used\cite{ueta}. Here we assume a
vector potential corresponding to the symmetric gauge (i.e.,
$\bbox{A}=\bbox{B}\times\bbox{r}/2$)
\begin{equation}
  \label{eq:mf-4}
  G\left(\bbox{r},\bbox{r'};E_n\right) \;=\; e^{i\left(x'y-y'x\right)/2\ell}
  \left(-\frac{m}{2\pi\hbar^2}\right)
  \Gamma\left(\frac{1}{2}-\epsilon\right) e^{-z/2}\,
  U\!\left(\frac{1}{2}-\epsilon,1,z\right)\;,  
\end{equation}
$\ell = \sqrt{\hbar/m\omega}$ is the so-called magnetic length, $\omega =
qB/m$ is the cyclotron frequency, $\varepsilon = E/\hbar\omega$,
$z=(\bbox{r}-\bbox{r'})^2/2\ell^2$, $\Gamma(x)$ is the Gamma
function\cite{a&s} and $U(a,b,x)$ is the logarithmic confluent
hypergeometric function\cite{a&s}. The derivation of Eq.(\ref{eq:mf-4}) is
beyond the scope of this article; the reader is referred to
Ref.~\onlinecite{ueta}.

The above Green's function can be evaluated numerically by employing the
double precision {\em SLATEC\/} subroutines\cite{slatec} {\tt DGAMMA}, for
the function $\Gamma(x)$, and {\tt DCHU}, for $U(a,b,x)$. Since the
evaluation of the Green's function and its normal derivative (which can be
expressed analytically) is very time consuming in the presence of a magnetic
field, it is recommended to apply Method III for determining the energy
spectrum of the system.

%---------------------------------------------------------------
\typeout{Conclusion}
\section{Conclusion}
\label{sec:conc}

In this article we have attempted to provide a self-contained, tutorial like
introduction to the {\em Boundary Integral Method\/} for calculating single
particle energy spectra in two-dimensional nano-devices. The BIM is suitable
whenever (i) a Green's function $G$ is available analytically and (ii) the
boundary condition at the boundary of the device is fairly simple. The
method applies to arbitrary shapes of the boundary. 

As we have shown, the BIM can be successfully applied to calculate the
energy spectrum of quantum billiards, allowing one to investigate the
quantum signatures of chaos in these systems. The numerical accuracy of the
BIM strongly depends on the degree of discretization of the billiard
boundary.  Unfortunately, by increasing the number of discretization points
along the billiard boundary, the needed computational resources seem to
increase more rapidly than the accuracy of the calculated energy levels.
Since the number of the needed discretization points along the billiard
boundary scales linearly with the cutoff wave vector $K_c$ [see
Eq.(\ref{eq:b16})], one can conclude that, in fact, the BIM allows one to
calculate the lowest few hundred energy levels of any quantum billiard. The
determination of higher energy levels, in general, becomes computationally
too expensive. Needles to say, the other existing numerical methods for
solving the Schr\"odinger equation present similar or even more stringent
limitations and altogether they perform worse than the BIM.

In conclusion, we would like to mention a few experimental confirmations of
the energy level fluctuations of quantum billiards described in this
article.  The revived interest in studying billiard spectra, in the context
of quantum chaos, has resulted in beautiful microwave
experiments\cite{stockmann90,graf92} designed to test the theoretical
predictions, based mainly on random matrix theories. These experiments
exploit the analogy between the Schr\"odinger wave equation of a quantum
particle in an infinite two-dimensional potential well and the Helmholtz
equation of the electromagnetic field in a resonant cavity. Thus, by
microwave measurements in the range of 0-25 GHz frequency in quasi
two-dimensional cavities shaped, e.g., in the form of a quarter stadium
billiard, up to few thousands eigenfrequencies were measured in
Refs.~\onlinecite{stockmann90,graf92}, and found in agreement with spectra
obtained by employing the BIM.  
Microwave measurements\cite{sridhar91,stein95} resulted also in the direct
observation of the eigenfunctions in microwave cavities of different shapes;
the eigenfunctions were also found to be in agreement with descriptions by
means of the BIM. A very recent microwave (``photon'') billiard
measurement\cite{stein95a} allowed for the first time the direct
experimental study of the energy level statistics in the presence of broken
time reversal symmetry; the level spacing distribution was found to conform
to a GUE form.

%----------------------------------------------------------------
\acknowledgments

We thank Professor S.-J.\ Chang, P.M.\ Goldbart and D.L.\ Maslov for useful
discussions.
This work was supported by the University of Illinois at Urbana-Champaign
and in part by the National Science Foundation Grant DMR91-20000 (through
STCS).

%----------------------------------------------------------------
\appendix
%______________________________________________________________
\typeout{Appendix: Free Particle Green's Function in 2D}
\section{Free Particle Green's Function in 2D}
\label{sec:free-particle}
In this appendix we derive the expression of the free particle Green's
function $G\left(\bbox{r},\bbox{r'};E\right)$ in two spatial dimensions. The
corresponding expression in $d$-dimensions can be obtained in a similar
fashion. 

The free particle Green's function is defined as the solution of the
equation [cf.\ Eq.(\ref{eq:bim3})]
\[
 \left(E+\frac{\hbar^2}{2m}\nabla_r^2\right)
 G\left(\bbox{r},\bbox{r'};E\right) \;=\;
 \delta\left(\bbox{r}-\bbox{r'}\right)\;, 
\]
or
\begin{equation}
\left(\nabla_r^2 + k^2\right) G\left(\bbox{r},\bbox{r'};k\right) \;=\;
\frac{2m}{\hbar^2}\delta\left(\bbox{r}-\bbox{r'}\right)\;,
  \label{eq:ap1-1}
\end{equation}
where $k\equiv\sqrt{2E/m\hbar^2}$ is the wave vector of the particle of
energy $E$, and we have replaced the energy variable in the Green's function
with $k$, i.e., $G(k)\equiv G(E)$. By changing variables
$\bbox{R}=\bbox{r}-\bbox{r'}$, which is equivalent to moving the origin of
the coordinate system to the point $\bbox{r'}$, the above equation becomes
\begin{equation}
  \left(\nabla_R^2 + k^2\right) G(\bbox{R};k) \;=\;
  \frac{2m}{\hbar^2}\delta(\bbox{R})\;,
  \label{eq:ap1-2}
\end{equation}
where $G(\bbox{R};k)\equiv G(\bbox{R},0;k)$. The fact that
Eq.(\ref{eq:ap1-2}) does not contain $\bbox{r'}$ and depends only on
$\bbox{R}$ is the consequence of translational symmetry.

One can solve (\ref{eq:ap1-2}) by of Fourier transform.  Inserting the
Fourier representations
\begin{equation}
  G(\bbox{R};k) \;=\; \int\frac{d^2\bbox{q}}{(2\pi)^2} \tilde{G}(\bbox{q};k)
  \exp(i\bbox{q}\bbox{R})\;,
  \label{eq:ap1-3}
\end{equation}
and
\begin{equation}
  \delta(\bbox{R}) \;=\; \int\frac{d^2\bbox{q}}{(2\pi)^2}
  \exp(i\bbox{q}\bbox{R})
  \label{eq:ap1-4}
\end{equation}
in Eq.(\ref{eq:ap1-2}), and identifying the Fourier coefficients on both
sides of the resulting equation, one arrives at
\begin{equation}
(-q^2+k^2)\tilde{G}(\bbox{q};k) \;=\; \frac{2m}{\hbar^2}\;.
  \label{eq:ap1-5}
\end{equation}
Inserting $\tilde{G}$ from (\ref{eq:ap1-5}) into (\ref{eq:ap1-3}) results in

\begin{equation}
  G(\bbox{R};k) \;=\; \frac{2m}{\hbar^2}\int\frac{d^2\bbox{q}}{(2\pi)^2}
  \frac{\exp(i\bbox{q}\bbox{R})}{k^2-q^2}\;.
  \label{eq:ap1-6}
\end{equation}
The two-dimensional integral is evaluated by using polar coordinates
$\bbox{q}=(q,\theta)$ as follows
\begin{equation}
  G(\bbox{R};k) \;=\; \frac{m}{\pi\hbar^2}\int_0^{\infty}\frac{q{dq}}{k^2-q^2}
  \frac{1}{2\pi}\int_0^{2\pi}{d\theta} \exp(iqR\cos\theta)\;.
  \label{eq:ap1-7}
\end{equation}
The second integral on the right hand side is identified as one of the
integral representations of the 0-th order Bessel function $J_0(qR)$ [cf.\
formula 8.4111.\ in Ref.\onlinecite{r&g}] and one obtains
\begin{equation}
  G(\bbox{R};k) \;=\; -\frac{m}{\pi\hbar^2}\int_0^{\infty}{dq} 
    \frac{q J_0(qR)}{q^2-k^2}\;.
  \label{eq:ap1-8}
\end{equation}

The integral on the RHS of (\ref{eq:ap1-8}) is ill defined due to the
singularity of the integrand at $q=\pm k$. However, the integral can be
regularized by adding to $k$ an infinitely small, positive imaginary part,
i.e., $k\rightarrow k+i\varepsilon$. In this case $k^2\rightarrow
(k+i\varepsilon)^2 = -(\varepsilon-ik)^2$, and according to the formula
6.5324 of Ref.\onlinecite{r&g} the integral in (\ref{eq:ap1-8}) is equal to
$K_0((\varepsilon-ik)R)$, where $K_0$ is the MacDonald (modified Bessel)
function, which is finite as $R\rightarrow\infty$. After taking the limit
$\varepsilon\rightarrow 0^+$, one obtains then
\begin{equation}
  G(\bbox{R};k) \;=\; -\frac{m}{\pi\hbar^2} K_0(-ikR)\;.
  \label{eq:ap1-9}
\end{equation}
Note that for $\varepsilon<0$ the above integral would be divergent for
$R\rightarrow\infty$. However, as long as we are not concerned with the
$R\rightarrow\infty$ behavior of $G(\bbox{R};k)$, the infinitesimal
$\varepsilon$ can be chosen either positive or negative. The choice
$\varepsilon>0$ is equivalent to the so-called {\em Sommerfeld radiation
  condition}\cite{jackson}.

Finally, by using the identity $K_0(z) = (i\pi/2)H_0^{(1)}(iz)$ [formula
8.4071 in Ref.\onlinecite{r&g}], where $z$ is an arbitrary complex number
and $H_0^{(1)}$ is the 0-th order Hankel function of the first kind, one
arrives at the expression of the free particle Green's function given by
(\ref{eq:b1}).

%______________________________________________________________
\typeout{Appendix: Evaluation of the singular integrals in the BIE}
\section{Evaluation of the singular integrals in the Boundary Integral
  Equation }
\label{sec:singular-int}

In order to calculate the LHS of Eq.(\ref{eq:bim9}), consider
first the case when the potential energy $V(\bbox{r})$ is zero and,
therefore, the relevant Green's function is given by
(\ref{eq:b1}). For $\varepsilon\equiv|\bbox{r}-\bbox{r'}|\rightarrow 0$
one can replace the Hankel function in the above equation by its limiting
form for small arguments\cite{H-lim-0}
\begin{equation}
  \label{eq:ap2-2}
  G\left(\bbox{r},\bbox{r'};E_n\right) \;\sim\;
  -\frac{m}{\pi\hbar^2}\ln(k\varepsilon)\;,\qquad\varepsilon\rightarrow 0\;.
\end{equation}
Next, we parameterize the arc of circle $C_{\varepsilon}$ through the angle
$\varphi$ (see Fig.\ref{fig:singular}) formed by the tangent AT$_1$ to
$\Gamma$ at A$\in\Gamma$ (of position vector $\bbox{r'}$) and the vector
$\bbox{\varepsilon}$. The angle $\varphi$ assumes values between zero and
$\theta(\bbox{r'})=\widehat{\text{T}_1\text{AT}_2}$, i.e., the exterior
angle made by the two tangents to $\Gamma$ at A. If the contour $\Gamma$ is
smooth then the tangents coincide and $\theta(\bbox{r'})=\pi$.  The arc
element along $C_{\varepsilon}$ is $ds(\bbox{r})=\varepsilon d\varphi$.
Since both $\psi_n$ and $\partial_{\nu}\psi_n$ are finite, in the limit
$\varepsilon\rightarrow 0$ the quantities can be replaced in (\ref{eq:bim9})
by their values at $\bbox{r}=\bbox{r'}$; one obtains then
\begin{equation}
  \label{eq:ap2-3}
  \lim_{\varepsilon\rightarrow 0}\frac{\hbar^2}{2m}\int_{C_{\varepsilon}}\! 
  ds(\bbox{r})
  G\left(\bbox{r},\bbox{r'};E_n\right)\partial_{\nu}\psi_n(\bbox{r}) \;=\;
  -\frac{1}{2\pi}\lim_{\varepsilon\rightarrow 0}
  \left[\varepsilon\ln(k\varepsilon)\right] \theta(\bbox{r'})
  \partial_{\nu}\psi_n(\bbox{r'}) \;=\; 0\;.
\end{equation}

The integral containing $\partial_{\nu}G$ in (\ref{eq:bim9}) can be
calculated in a similar fashion. According to Eqs.(\ref{eq:bim8}) and
(\ref{eq:b1}) one can write successively
\begin{eqnarray}
  \label{eq:ap2-4}
  \partial_{\nu}G\left(\bbox{r},\bbox{r'};E_n\right) &\;=\;&
  \bbox{\nu}(\bbox{r})\cdot\nabla_r \left[-\frac{im}{2\hbar^2}
  H_0^{(1)}\left(k|\bbox{r}-\bbox{r'}|\right)\right]\\
  &\;=\;& \frac{imk}{2\hbar^2}\:\left[\bbox{\nu}(\bbox{r})\cdot
  \frac{\bbox{r}-\bbox{r'}}{|\bbox{r}-\bbox{r'}|}\right]\:
  H_{1}^{(1)}\left(k|\bbox{r}-\bbox{r'}|\right)\;,\nonumber
\end{eqnarray}
where we used $dH_{0}^{(1)}(z)/dz = -H_{1}^{(1)}(z)$ [cf.\ formula 9.1.30 in
Ref.\onlinecite{a&s}]. On the arc of circle $C_{\varepsilon}$ the dot
product in (\ref{eq:ap2-4}) is equal to one (see Fig.\ref{fig:singular});
taking into account the limiting form of $H_1^{(1)}$ for small
arguments\cite{H-lim-0}, one can write
\begin{eqnarray}
  \label{eq:ap2-5}
  \lim_{\varepsilon\rightarrow 0}\frac{\hbar^2}{2m}
  \int_{C_{\varepsilon}}\!ds(\bbox{r}) \psi_n(\bbox{r})
  \partial_{\nu}G\left(\bbox{r},\bbox{r'};E_n\right) &\;=\;&
  \frac{\hbar^2}{2m}\psi_n(\bbox{r'}) \lim_{\varepsilon\rightarrow 0} 
  \int_0^{\theta(\bbox{r'})}\varepsilon d\varphi 
  \left(\frac{imk}{2\hbar^2}\right)
  \left(-\frac{2i}{\pi k\varepsilon}\right)\\
  &\;=\;& \frac{\theta(\bbox{r'})}{2\pi}\psi_n(\bbox{r'})\nonumber\;.
\end{eqnarray}
For a smooth boundary $\Gamma$, where $\theta(\bbox{r'})=\pi$,
Eqs.(\ref{eq:ap2-3}-\ref{eq:ap2-5}) provide the result given in
(\ref{eq:bim9}).

For a finite potential energy $V(\bbox{r})$, in general, there is no simple
analytical expression for the Green's function $G$ and the validity of
Eq.(\ref{eq:bim9}) is questionable. However, by assuming on physical grounds
that $V(\bbox{r})$ is finite for all $\bbox{r}\in{\cal D}$, one can realize
that the result (\ref{eq:bim9}) holds in this case too. Indeed, when
$\varepsilon$ is small the potential energy is almost constant in the
vicinity of $\bbox{r'}$ (point A in Fig.\ref{fig:singular}) and, therefore,
one can approximate the Green's function with the corresponding expression
valid for a constant $V\equiv V(\bbox{r'})$. The approximation becomes exact
in the limit $\varepsilon\rightarrow 0$. But $G$ for a constant potential
energy has essentially the same form as for a free particle
[Eq.(\ref{eq:b1})] and, therefore, it has the same type of logarithmic
singularity at $\bbox{r}=\bbox{r'}$. Since the actual value of the integral
(\ref{eq:bim9}) is determined solely by the type of this singularity of the
Green's function one may conclude that the result derived in this appendix
holds in general.
%______________________________________________________________
\typeout{Appendix: Analytical Solution of the BIE for a CB}
\section{Analytical Solution of the Boundary Integral Equation for a
  Circular Billiard} 
\label{sec:circle}

In this appendix we solve analytically the BIE (\ref{eq:b6}) for a circular
billiard of unit radius. For the unit circle ${\cal L}=2\pi$ and, according
to Eq.(\ref{eq:b7c}), one finds $K_j=j$, with $j=0,\pm 1,\ldots$. By using
the Fourier representation (\ref{eq:b7a}) for $u(s)$, the BIE becomes
\begin{equation}
  \label{eq:ap3-1}
  \sum_{j=-\infty}^{\infty} u_j \int_0^{2\pi} \!ds\,
  G\left(s,s';k\right)\exp(i js)\;=\;0\;.
\end{equation}
The expression (\ref{eq:ap1-6}) of the free particle Green's function, in
the present case, can be written 
\begin{eqnarray}
  \label{eq:ap3-2}
  G\left(s,s';k\right) &\;=\;& \frac{2m}{\hbar^2}
  \int\frac{d^2\bbox{q}}{(2\pi)^2}
  \frac{\exp\{i\bbox{q}[\bbox{r}(s)-\bbox{r}(s')]\}}{k^2-q^2}\\
  &\;=\;& \frac{m}{\pi\hbar^2}\int_0^{\infty}\frac{qdq}{k^2-q^2}
  \frac{1}{2\pi}\int_0^{2\pi}d\theta\exp[iq\cos(s-\theta)]
  \exp[-iq\cos(s'-\theta)]%
  \nonumber\;,
\end{eqnarray}
where $q$ and $\theta$ are the polar coordinates of the 2D vector
$\bbox{q}$. Inserting (\ref{eq:ap3-2}) in (\ref{eq:ap3-1}) one obtains (the
irrelevant constant factor $2m/\hbar^2$ can be dropped)
\begin{eqnarray}
  \label{eq:ap3-3}
  \sum_{j=-\infty}^{\infty} u_j \int_0^{\infty}\frac{qdq}{k^2-q^2} 
  \frac{1}{2\pi}\int_0^{2\pi}d\theta \exp[-iq\cos(s'-\theta)]& &\\
  \times\frac{1}{2\pi}\int_0^{2\pi}ds 
  \exp[iq\cos(s-\theta)]\exp(i js) &\;=\;& 0\;.\nonumber
\end{eqnarray}
By taking into account the integral representation of the integer Bessel
function [formula 8.4111.\ in Ref.\onlinecite{r&g}], the integral over $s$
in (\ref{eq:ap3-3}) can be evaluated exactly as follows
\begin{eqnarray}
  \label{eq:ap3-4}
  \frac{1}{2\pi}\int_0^{2\pi}ds \exp[iq\cos(s-\theta)]\exp(i js) 
  &\;=\;& \left[\frac{1}{2\pi}\int_0^{2\pi}ds%
  \exp\{i[q\cos(s-\theta)+j(s-\theta)]\}\right]\\
  \times\exp(i j\theta) &\;=\;& i^j J_j(q) \exp(i j\theta)\nonumber\;.
\end{eqnarray}
Similarly, the integral with respect to $\theta$ in (\ref{eq:ap3-3}) gives
\begin{equation}
  \label{eq:ap3-5}
  \frac{1}{2\pi}\int_0^{2\pi}d\theta \exp[-iq\cos(s'-\theta)]\exp(i j\theta)
  \;=\; (-i)^j J_j(q) \exp(i js')\;.
\end{equation}
With the last two results, Eq.(\ref{eq:ap3-3}) becomes
\begin{equation}
  \label{eq:ap3-6}
  \sum_{j=-\infty}^{\infty} u_j \exp(i js')\int_0^{\infty}dq
  \frac{q\left[J_j(q)\right]^2}{k^2-q^2}  \;=\; 0\;.
\end{equation}
By employing formulas 6.535, 8.4061 and 8.4071 in Ref.\onlinecite{r&g}, the
integral with respect to $q$ can also be calculated exactly as follows
\begin{equation}
  \label{eq:ap3-7}
  \int_0^{\infty}dq\frac{q\left[J_j(q)\right]^2}{k^2-q^2}      \;=\;
  - \int_0^{\infty}dq\frac{q\left[J_j(q)\right]^2}{q^2+(ik)^2} \;=\;
  -I_j(-ik) K_j(-ik) \;=\; -\frac{\pi}{2}J_j(k) H_j^{(1)}(k)\;.
\end{equation}
Here $I_j$ and $K_j$ are imaginary argument Bessel functions. 

Finally, the BIE for the circle billiard of unit radius can be written as
\begin{equation}
  \label{eq:ap3-8}
  \sum_{j=-\infty}^{\infty} u_j \exp(i js') J_j(k) H_j^{(1)}(k) \;=\; 0\;,
\end{equation}
Since $\exp(i js)$, $j=0,\pm 1\ldots$, form a complete orthonormal set,
the above equation tells us that the expansion coefficients should be all
equal to zero. The eigenenergies correspond to those $k$ values for which
non trivial $u_j$'s exist. Recalling that the Hankel functions have no real
roots, one obtains the same eigenenergy equation (\ref{eq:b10}) as in
Sec.\ref{sec:methods}.

By taking the Fourier transform of Eq.(\ref{eq:ap3-8}) with respect to
$s'$, one can see that the matrix $A_{ij}(k)$ defined by (\ref{eq:b8a}) is
indeed diagonal and
\begin{equation}
  \label{eq:ap3-9}
  A_{ij}(k) \;\propto\; J_j(k) H_j^{(1)}(k)\:\delta_{ij}\;.
\end{equation}

%______________________________________________________________
\typeout{Superposition of $N$ Independent GOE Distributions}
\section{Superposition of $N$ Independent Spectra with GOE Level Spacing
  Distribution }
\label{sec:sup-GOE}

In this Appendix we derive the level spacing distribution function
$P^{(N)}(s)$ corresponding to the superposition of $N$ independent spectra
with GOE level spacing distribution.
In general, $P(s)$ can be expressed\cite{haake}

\begin{equation}
  \label{eq:sup-GOE-1}
  P(s) \;=\; \frac{\partial^2{\cal E}(s)}{\partial\,s^2}\;,
\end{equation}
where ${\cal E}(s)$, the so-called {\em gap probability}, gives the
probability that the energy interval $(E,E+s)$ lacks energy levels. 

Let us consider $N$ independent (i.e., uncorrelated) sets of energy levels
 with GOE level spacing distribution 

\begin{equation}
  \label{eq:sup-GOE-2}
  P_i(s) \;=\; \frac{\pi}{2}\frac{s}{N^2}\,\exp\left[-\frac{\pi}{4}
    \left(\frac{s}{N}\right)^2 \right]\;, \qquad i\:=\:1,\ldots,N
\end{equation}
The probability density (\ref{eq:sup-GOE-2}) is normalized as follows
\begin{equation}
  \label{eq:sup-GOE-3}
  \int_0^{\infty}\!ds\,P_i(s)\;=\;1\;, \qquad
  \langle s_i\rangle \;\equiv\; \int_0^{\infty}\!ds\:s\,P_i(s)\;=\;N\;.
\end{equation}
Note that the choice $\langle s_i\rangle=N$ for each set of levels leads to
a unit mean level spacing $\langle s^{(N)} \rangle$ for the spectrum
comprising all $N$ energy spectra.

According to Eqs.(\ref{eq:sup-GOE-1})-(\ref{eq:sup-GOE-2}), the
individual gap probabilities can be expressed as
\begin{eqnarray}
  \label{eq:sup-GOE-4}
  {\cal E}_i(s) &\;=\;& \frac{1}{N}\int_s^{\infty}\!dx
  \int_x^{\infty}\!dy\,P_i(y)\\
  &\;=\;& \frac{2}{\sqrt{\pi}}\int_{\frac{\sqrt{\pi}\,s}{2N}}^{\infty}\!dt\,
    \exp\left(-t^2\right) \;=\;
    \text{erfc}\left(\frac{\sqrt{\pi}\,s}{2N}\right) \;,\nonumber 
\end{eqnarray}
where erfc$(z)$ is the complementary error function\cite{a&s}.
Since the energy spectra are uncorrelated, the gap probability of the
combined spectrum is given by
\begin{equation}
  \label{eq:sup-GOE-5}
  {\cal E}^{(N)}(s) \;=\; \prod_{i=1}^N {\cal E}_i(s) \;=\;
  \left[\text{erfc}\left(\frac{\sqrt{\pi}}{2}\frac{s}{N}\right)\right]^N\;,
\end{equation}
and, according to (\ref{eq:sup-GOE-1}), the desired level spacing
distribution function becomes
\begin{equation}
  \label{eq:sup-GOE-6}
  P^{(N)}(s) \;=\; \frac{\partial^2}{\partial\,s^2}
  \left[\text{erfc}\left(\frac{\sqrt{\pi}}{2}\frac{s}{N}\right)\right]^N\;.
\end{equation}

Note that the above method of calculating $P(s)$ is rather general and
applies also when the independent spectra have arbitrary statistics. For
example, one could calculate the level spacing distribution of the
superposition of an arbitrary number of spectra, some of them obeying
Poisson statistics and the rest GOE statistics. For further details the
reader is referred to Refs.~\onlinecite{bohigas89,mehta}.

%______________________________________________________________
%\typeout{Appendix: Green's Function of a Charged Particle in Magnetic Field}
%\section{Green's Function of a Charged Particle in Magnetic Field}
%\label{sec:gf-mag}
%----------------------------------------------------------------
\typeout{References}

%------------------------------------------------------------------

\begin{references}
\bibitem{H-lim-0} For $z\rightarrow 0$ one has $H_0^{(1)}(z)\sim (2i/\pi)
  \ln(z)$, and $H_1^{(1)}(z)\sim -(2i/\pi{z})$; see, e.g., formulas 9.1.8-9
  in Ref.\protect\onlinecite{a&s}
\end{references}
\end{document}